\documentclass[12pt, draftcls, onecolumn]{IEEEtran}

\usepackage{setspace}
\usepackage{amsmath}
\usepackage{amssymb}
\usepackage[dvips]{graphicx}
\usepackage{epsfig}

\newcommand{\x}{\mathbf{x}}
\newcommand{\y}{\mathbf{y}}

\newcommand{\z}{\mathbf{z}}

\newcommand{\he}{\hat{h}}

\newcommand{\figsize}{0.65}

\newcommand{\tsnr}{{\text{\footnotesize{SNR}}}}

\newtheorem{prop:asympcap}{Theorem}
\newtheorem{prop:flashminbitenergy}[prop:asympcap]{Theorem}
\newtheorem{prop:flashbitenergy}[prop:asympcap]{Theorem}
\newtheorem{prop:pasympcap}[prop:asympcap]{Theorem}
\newtheorem{lemma:bitenergylow}[prop:asympcap]{Theorem}

\begin{document}


\title{Achievable Rates and Optimal Resource Allocation for Imperfectly-Known Fading Relay Channels}



%
{\small{
\author{\authorblockN{Junwei Zhang, \and Mustafa Cenk Gursoy}
\\
\authorblockA{Department of Electrical Engineering\\
University of Nebraska-Lincoln, Lincoln, NE 68588\\ Email:
jzhang13@bigred.unl.edu, gursoy@engr.unl.edu}}}}


\maketitle


\begin{abstract}\footnote{This work was supported in part by the NSF CAREER Grant CCF-0546384.
The material in this paper was presented in part at the 45th Annual
Allerton Conference on Communication, Control and Computing in Sept.
2007.} In this paper, achievable rates and optimal resource
allocation strategies for imperfectly-known fading relay channels
are studied. It is assumed that communication starts with the
network training phase in which the receivers estimate the fading
coefficients of their respective channels. In the data transmission
phase, amplify-and-forward and decode-and-forward relaying schemes
with different degrees of cooperation are considered, and the
corresponding achievable rate expressions are obtained. Three
resource allocation problems are addressed: 1) power allocation
between data and training symbols; 2) time/bandwidth allocation to
the relay; 3) power allocation between the source and relay in the
presence of total power constraints. The achievable rate expressions
are employed to identify the optimal resource allocation strategies.
Finally, energy efficiency is investigated by studying the bit
energy requirements in the low-$\tsnr$ regime.


\end{abstract}

\begin{spacing}{1.63}

\section{Introduction}

In wireless communications, deterioration in  performance is
experienced due to various impediments such as interference,
fluctuations in power due to reflections and attenuation, and
randomly-varying channel conditions caused by mobility and changing
environment. Recently, cooperative wireless communications has
attracted much interest as a technique that can mitigate these
degradations and provide higher rates or improve the reliability
through diversity gains. The relay channel was first introduced by
van der Meulen in \cite{meulen}, and initial research was primarily
conducted to understand the rates achieved in relay channels
\cite{cover}, \cite{Gamal}. More recently, diversity gains of
cooperative transmission techniques have been studied in
\cite{sendonaris1}--\cite{jnlbook}. In \cite{jnl}, several
cooperative protocols have been proposed, with
amplify-and-forward~(AF) and decode-and-forward~(DF) being the two
basic relaying schemes. The performance of these protocols are
characterized in terms of outage events and outage probabilities.
In~\cite{Nabar}, three different time-division AF and DF cooperative
protocols  with different the degrees of broadcasting and receive
collision are studied. In general, the area has seen an explosive
growth in the number of studies (see e.g., \cite{host-madsen1},
\cite{kramer}, \cite{Liang}, \cite{Mitran}, \cite{Yao}, and
references therein). An excellent review of cooperative strategies
from both rate and diversity improvement perspectives is provided in
\cite{tutorial} in which the impacts of cooperative schemes on
device architecture and higher-layer wireless networking protocols
are also addressed. Recently, a special issue has been dedicated to
models, theory, and codes for relaying and cooperation in
communication networks in \cite{specialissue}.

As noted above, studies on relaying and cooperation are numerous.
However, most work has assumed that the channel conditions are
perfectly known at the receiver and/or transmitter sides. Especially
in mobile applications, this assumption is unwarranted as
randomly-varying channel conditions can be learned by the receivers
only imperfectly. Moreover, the performance analysis of cooperative
schemes in such scenarios is especially interesting and called for
because relaying introduces additional channels and hence increases
uncertainty in the model if the channels are known only imperfectly.
Recently, Wang \emph{et al.} in \cite{wang} considered
pilot-assisted transmission over wireless sensory relay networks,
and analyzed scaling laws achieved by the amplify-and-forward scheme
in the asymptotic regimes of large nodes, large block length, and
small $\tsnr$ values. In this study, the channel conditions are
being learned only by the relay nodes. In \cite{tse}, Avestimehr and
Tse studied the outage capacity of slow fading relay channels. They
showed that Bursty Amplify-Forward strategy achieves the outage
capacity in the low SNR and low outage probability regime.
Interestingly, they further proved that the optimality of Bursty AF
is preserved even if the receivers do not have prior knowledge of
the channels.

In this paper, we study the achievable rates of imperfectly-known
fading relay channels. We assume that transmission takes place in
two phases: network training phase and data transmission phase. In
the network training phase, a priori unknown fading coefficients are
estimated at the receivers with the assistance of pilot symbols.
Following the training phase, AF and DF relaying techniques with
different degrees of cooperation are employed in the data
transmission. We first obtain achievable rate expressions for
different relaying protocols and subsequently identify optimal
resource allocation strategies that maximize the rates. We consider
three types of resource allocation problems: 1) power allocation
between data and training symbols; 2) time/bandwidth allocation to
the relay; 3) power allocation between the source and relay if there
is a total power constraint in the system. Finally, we investigate
the energy efficiency by finding the bit energy requirements in the
low-$\tsnr$ regime.

The organization of the rest of the paper is as follows. In Section
\ref{sec:model}, we describe the channel model. Network training and
data transmission phases are explained in Section
\ref{sec:training}. We obtain the achievable rate expressions in
Section \ref{sec:achievablerates} and study the optimal resource
allocation strategies in Section \ref{sec:allocation}. We discuss
the energy efficiency in the low-$\tsnr$ regime in Section
\ref{sec:energyefficiency}. Finally, we provide conclusions in
Section \ref{sec:conclusion}.

\section{Channel Model} \label{sec:model}

We consider the three-node relay network which consists of a source,
destination, and a relay node. \vspace{-1cm}
\centerline{ \setlength{\unitlength}{0.011in}
\begin{picture}(275,175)(105,632)
\thicklines \put(140,720){\circle*{10}} \put(240,770){\circle*{10}}
\put(340,720){\circle*{10}} \put(150,725){\vector(2,1){80}}
\put(250,765){\vector(2,-1){80}} \put(150,720){\vector( 1,0){180}}
\put(117,690){\makebox(0,0)[lb]{\raisebox{0pt}[0pt][0pt]{Source}}}
\put(220,785){\makebox(0,0)[lb]{\raisebox{0pt}[0pt][0pt]{Relay}}}
\put(300,690){\makebox(0,0)[lb]{\raisebox{0pt}[0pt][0pt]{Destination}}}
\put(170,755){\makebox(0,0)[lb]{\raisebox{0pt}[0pt][0pt]{$h_{sr}$}}}
\put(283,755){\makebox(0,0)[lb]{\raisebox{0pt}[0pt][0pt]{$h_{rd}$}}}
\put(230,705){\makebox(0,0)[lb]{\raisebox{0pt}[0pt][0pt]{$h_{sd}$}}}
\put(360,710){\makebox(0,0)[lb]{\raisebox{0pt}[0pt][0pt]{$\y_{d}$}}}
\put(360,730){\makebox(0,0)[lb]{\raisebox{0pt}[0pt][0pt]{$\y_{d}^r$}}}
\put(117,715){\makebox(0,0)[lb]{\raisebox{0pt}[0pt][0pt]{$\x_s$}}}
\put(215,773){\makebox(0,0)[lb]{\raisebox{0pt}[0pt][0pt]{$\y_r$}}}
\put(252,773){\makebox(0,0)[lb]{\raisebox{0pt}[0pt][0pt]{$\x_r$}}}
\end{picture} }
Source-destination, source-relay, and relay-destination channels are
modeled as  Rayleigh block-fading channels with fading coefficients
denoted by $h_{sr}$, $ h_{sd}$, and $h_{rd}$, respectively for each
channel. Due to the block-fading assumption, the fading coefficients
$h_{sr}\sim\mathcal {C}\mathcal {N}(0,{\sigma_{sr}}^2)$,
$h_{sd}\sim\mathcal {C}\mathcal {N}(0,{\sigma_{sd}}^2)$, and
$h_{rd}\sim\mathcal {C}\mathcal
{N}(0,{\sigma_{rd}}^2)$\footnote{$x\sim\mathcal {C}\mathcal
{N}(d,{\sigma^2)}$ is used to denote a proper complex Gaussian
random variable with mean $d$ and variance $\sigma^2$.} stay
constant for a block of $m$ symbols before they assume independent
realizations for the following block. In this system, the source
node tries to send information to the
destination node with the help of the intermediate relay node. 
It is assumed that the source, relay, and destination nodes do not
have prior knowledge of the realizations of the fading coefficients.
The transmission is conducted in two phases: network training phase
in which the fading coefficients are estimated at the receivers, and
data transmission phase. Overall, the source and relay are subject
to the following power constraints in one block: \vspace{-0.45cm}
\begin{equation} \label{spower}
\|{\mathbf{x}_{s,t}}\|^2+E\{\|{\mathbf{x}_s}\|^2\}\leq mP_{s},
\end{equation}
\begin{equation} \label{rpower}
\|{\mathbf{x}_{r,t}}\|^2+E\{\|{\mathbf{x}_r}\|^2\}\leq mP_{r}.
\end{equation}
where $\mathbf{x}_{s,t}$ and $\mathbf{x}_{r,t}$ are the source and
relay training signal vectors, respectively, and $\mathbf{x}_{s}$
and $\mathbf{x}_{r}$ are the corresponding source and relay data
vectors.

\section{Network Training and Data Transmission}
\label{sec:training}

\subsection{Network Training Phase}

Each block transmission starts with the training phase. In the first
symbol period, source transmits a pilot symbol to enable the relay
and destination to estimate the channel coefficients $h_{sr}$ and
$h_{sd}$, respectively. In the average power limited case, sending a
single pilot is optimal because instead of increasing the number of
pilot symbols, a single pilot with higher power can be used. The
signals received by the relay and destination are \vspace{-0.3cm}
\begin{equation} \label{srtraining}
y_{r,t}=h_{sr}x_{s,t}+n_r, \quad \text{and} \quad 
y_{d,t}=h_{sd}x_{s,t}+n_d,
\end{equation}
respectively. Similarly, in the second symbol period, relay
transmits a pilot symbol to enable the destination to estimate the
channel coefficient $h_{rd}$. The signal received by the destination
is
\begin{equation} \label{rdtraining}
y_{d,t}^{r}=h_{rd}x_{r,t}+n_d^r.
\end{equation}
In the above formulations, $n_r\sim\mathcal {C}\mathcal {N}(0,N_0)$,
$n_d\sim\mathcal {C}\mathcal {N}(0,N_0)$, and $n_d^r\sim\mathcal
{C}\mathcal {N}(0,N_0)$ represent independent Gaussian noise samples
at the relay  and the destination nodes.

In the training process, it is assumed that the receivers employ
minimum mean-square-error (MMSE) estimation. We assume that the
source allocates $\delta_s$ of its total power for training while
the relay allocates $\delta_r$ of its total power for training. As
described in  \cite{gursoy}, the MMSE estimate of $h_{sr}$ is given
by

\begin{equation}\label{est}
\hat{h}_{sr}=\frac{\sigma_{sr}^2\sqrt{\delta_smP_s}}{\sigma_{sr}^2\delta_smP_s+N_0}y_{r,t}
,
\end{equation}
where $y_{r,t}\sim\mathcal {C}\mathcal
{N}(0,\sigma_{sr}^2\delta_smP_s+N_0)$. We denote by
 $\tilde{h}_{sr}$  the estimate error which is a zero-mean complex
Gaussian random variable with variance $
var(\tilde{h}_{sr})=\frac{\sigma_{sr}^2N_0}{\sigma_{sr}^2\delta_smP_s+N_0}.
$ Similarly, for the fading coefficients $h_{sd}$ and $h_{rd}$, we
have
\begin{gather}
\hat{h}_{sd}=\frac{\sigma_{sd}^2\sqrt{\delta_smP_s}}{\sigma_{sd}^2\delta_smP_s+N_0}y_{d,t},\quad
y_{d,t}\sim\mathcal {C}\mathcal
{N}(0,\sigma_{sd}^2\delta_smP_s+N_0), \quad
var(\tilde{h}_{sd})=\frac{\sigma_{sd}^2N_0}{\sigma_{sd}^2\delta_smP_s+N_0},
\label{eq:hsdest}
\\
\hat{h}_{rd}=\frac{\sigma_{rd}^2\sqrt{\delta_rmP_r}}{\sigma_{rd}^2\delta_rmP_r+N_0}y_{d,t}^r,\quad
y_{d,t}^r\sim\mathcal {C}\mathcal
{N}(0,\sigma_{rd}^2\delta_rmP_r+N_0), \quad 
var(\tilde{h}_{rd})=\frac{\sigma_{rd}^2N_0}{\sigma_{rd}^2\delta_rmP_r+N_0}.
\label{eq:hrdest}
\end{gather}
With these estimates, the fading coefficients can now be expressed
as
\begin{gather}\label{ttt}
h_{sr}=\hat{h}_{sr} +\tilde{h}_{sr}, \quad
 h_{sd}=\hat{h}_{sd}+\tilde{h}_{sd},
\quad 
 h_{rd}=\hat{h}_{rd} +\tilde{h}_{rd}.
\end {gather}

\subsection {Data Transmission Phase}

The practical relay node usually cannot transmit and receive data
simultaneously. Thus, we assume that the relay works under
half-duplex constraint. Hence, the relay first listens and then
transmits. As discussed in the previous section, within a block of
$m$ symbols, the first two symbols are allocated to network
training. In the remaining duration of $m-2$ symbols, data
transmission takes place. We introduce the relay transmission
parameter $\alpha$ and assume that $\alpha(m-2)$ symbols are
allocated for relay transmission. Hence, $\alpha$ can be seen as the
fraction of total time or bandwidth allocated to the relay. Note
that the parameter $\alpha$ enables us to control the degree of
cooperation. We consider several transmission protocols which can be
classified into two categories by whether or not the source and
relay simultaneously transmits information: non-overlapped and
overlapped transmission. Note that in both cases, the relay
transmits over a duration of $\alpha(m-2)$ symbols. In
non-overlapped transmission, source transmits over a duration of
$(1-\alpha)(m-2)$ symbols and becomes silent as the relay transmits.
On the other hand, in overlapped transmission,  source transmits all
the time and sends $m-2$ symbols in each block.

We assume that the data vectors $\mathbf{x}_{s}$ and
$\mathbf{x}_{r}$ are composed of independent random variables with
equal energy. Hence, the covariance matrices of $\mathbf{x}_{s}$ are
given by
\begin{equation}\label{xsp2}
E\{\mathbf{x}_{s}\mathbf{x}_{s}^\dagger\}=P_{s1}'\,\,\mathbf{I}=\frac{(1-\delta_s)mP_s}{(m-2)(1-\alpha)}\,\,\mathbf{I},
\quad  \text{and} \quad
E\{\mathbf{x}_{s}\mathbf{x}_{s}^\dagger\}=P_{s2}'\,\,\mathbf{I}=\frac{(1-\delta_s)mP_s}{(m-2)}\,\,\mathbf{I},
\end{equation}
in non-overlapped and overlapped transmissions, respectively. The
covariance matrix for $\mathbf{x}_{r}$ is
\begin{equation}\label{xrp2}
E\{\mathbf{x}_{r}\mathbf{x}_{r}^\dagger\}=P_{r}'\,\,\mathbf{I}=\frac{(1-\delta_r)mP_r}{(m-2)\alpha}\,\,\mathbf{I}.
\end{equation}

\subsubsection{Non-overlapped transmission}
We first consider the two simplest cooperative protocols:
non-overlapped AF, and non-overlapped DF with repetition coding
where the relay decodes the message and re-encodes it using the same
codebook as the source. In these protocols, since the relay either
amplifies the received signal, or decodes it but uses the same
codebook as the source when forwarding, source and relay should be
allocated equal time slots in the cooperation
phase. 
Therefore, we initially have direct transmission from the source to
the destination without any aid from the relay over a duration of
$(1-2\alpha)(m-2)$ symbols. In this phase, source sends
$\mathbf{x}_{s1}$ and the received signal at the destination is
given by
\begin{equation}\label{sdtxddd}
\mathbf{y}_{d1}=h_{sd}\mathbf{x}_{s1}+\mathbf{n}_{d1}.
\end{equation}
Subsequently, cooperative transmission starts. At first, the source
transmits an $\alpha(m-2)$-dimensional symbol vector
$\mathbf{x}_{s2}$ which is received at the the relay and the
destination, respectively, as
\begin{equation}\label{srtx}
\mathbf{y}_r=h_{sr}\mathbf{x}_{s2}+\mathbf{n}_r, \quad \text{and}
\quad
\mathbf{y}_{d2}=h_{sd}\mathbf{x}_{s2}+\mathbf{n}_{d2}.
\end{equation}
For compact representation, we denote the signal received at the
destination directly from the source by $\y_d = [\y_{d1}^T \,\,
\y_{d2}^T]^T$ where $T$ denotes the transpose operation. Next, the
source becomes silent, and the relay transmits an
$\alpha(m-2)$-dimensional symbol vector $\mathbf{x}_r$ which is
generated from the previously received $\mathbf{y}_r$ \cite{jnl}
\cite{jnlbook}. This approach corresponds to protocol 2 in
\cite{Nabar}, which realizes the maximum degrees of broadcasting and
exhibits no receive collision. The destination receives
\begin{equation}\label{rdtx}
\mathbf{y}_d^r=h_{rd}\mathbf{x}_{r}+\mathbf{n}_d^r.
\end{equation}
After substituting the expressions in (\ref{ttt}) into
(\ref{sdtxddd})--(\ref{rdtx}), we have
\begin{gather}\label{sdtx1}
\mathbf{y}_{d1}=\hat{h}_{sd}\mathbf{x}_{s1}+\tilde{h}_{sd}\mathbf{x}_{s1}+\mathbf{n}_{d1},
\quad
\mathbf{y}_r=\hat{h}_{sr}\mathbf{x}_{s2}+\tilde{h}_{sr}\mathbf{x}_{s2}+\mathbf{n}_r,
\quad
\mathbf{y}_{d2}=\hat{h}_{sd}\mathbf{x}_{s2}+\tilde{h}_{sd}\mathbf{x}_{s2}+\mathbf{n}_{d2},
\\
\mathbf{y}_d^r=\hat{h}_{rd}\mathbf{x}_{r}+\tilde{h}_{rd}\mathbf{x}_{r}+\mathbf{n}_d^r.\label{rdtx1}
\end{gather}
We define the source data vector as $\x_s =
[\x_{s1}^T\,\,\x_{s2}^T]^T$. Note that we have $0 < \alpha \le 1/2$
for AF and repetition coding DF. Therefore, $\alpha = 1/2$ models
full cooperation while we have noncooperative communications as
$\alpha \to 0$. It should also be noted that $\alpha$ should in
general be chosen such that $\alpha(m-2)$ is an integer.

For non-overlapped transmission, we also consider DF with parallel
channel coding, in which the relay uses a different codebook to
encode the message. In this case, the source and relay do not have
to be allocated the same duration in the cooperation phase.
Therefore, source transmits over a duration of $(1-\alpha)(m-2)$
symbols while the relay transmits in the remaining duration of
$\alpha(m-2)$ symbols. Clearly, the range of $\alpha$ is now $0<
\alpha < 1$. In this case, the input-output relations are given by
(\ref{srtx}) and (\ref{rdtx}). Since there is no separate direct
transmission, $\x_{s2} = \x_s$ and $\y_{d2} = \y_d$ in (\ref{srtx}).
Moreover, the dimensions of the vectors $\mathbf{x}_{s},
\mathbf{y}_{d}, \mathbf{y}_r$ are now $(1-\alpha)(m-2)$, while
$\mathbf{x}_{r}$ and $\mathbf{y}_d^r$ are vectors of dimension
$\alpha(m-2)$.

\subsubsection{Overlapped transmission}
In this category, we consider a more general and complicated
scenario in which the source transmits all the time.  In AF and
repetition DF, similarly as in the non-overlapped model, cooperative
transmission takes place in the duration of $2\alpha(m-2)$ symbols.
The remaining duration of $(1-2\alpha)(m-2)$ symbols is allocated to
unaided direct transmission from the source to the destination.
Again, we have $0<\alpha \leq 1/2$ in this setting. In these
protocols, the input-output relations are expressed as follows:
\begin{align}\label{ovdtx}
\hspace{-.13cm}\mathbf{y}_{d1}=h_{sd}\mathbf{x}_{s1}+\mathbf{n}_{d1},
\quad
\mathbf{y}_r=h_{sr}\mathbf{x}_{s21}+\mathbf{n}_r, \quad
\mathbf{y}_{d2}=h_{sd}\mathbf{x}_{s21}+\mathbf{n}_{d2}, \quad
\text{and} \quad
\mathbf{y}_d^r=h_{sd}\mathbf{x}_{s22}+h_{rd}\mathbf{x}_{r}+\mathbf{n}_d^r.
\end{align}
Above, $\mathbf{x}_{s1},\mathbf{x}_{s21},\mathbf{x}_{s22}$, which
have respective dimensions $(1-2\alpha)(m-2)$, $\alpha(m-2)$ and
$\alpha(m-2)$, represent the source data vectors sent in direct
transmission, cooperative transmission when relay is listening, and
cooperative transmission when relay is transmitting, respectively.
Note again that the source transmits all the time. $\mathbf{x}_{r}$
is the relay's data vector with dimension $\alpha(m-2)$.
$\mathbf{y}_{d1},\mathbf{y}_{d2},\mathbf{y}_d^r$ are the
corresponding received vectors at the destination, and
$\mathbf{y}_r$ is the received vector at the relay. The input vector
$\mathbf{x}_s$ now is defined as
$\mathbf{x}_{s}=[\mathbf{x}_{s1}^T,\mathbf{x}_{s21}^T,\mathbf{x}_{s22}^T]^T$
and we again denote  $\y_d = [\y_{d1}^T \,\, \y_{d2}^T]^T$. If we
express the fading coefficients as $h = \hat{h} + \tilde{h}$ in
(\ref{ovdtx}), we obtain the following input-output relations:
\begin{gather}\label{srtxx1x}
\mathbf{y}_{d1}=\hat{h}_{sd}\mathbf{x}_{s1}+\tilde{h}_{sd}\mathbf{x}_{s1}+\mathbf{n}_{d1},
\quad
\mathbf{y}_r=\hat{h}_{sr}\mathbf{x}_{s21}+\tilde{h}_{sr}\mathbf{x}_{s21}+\mathbf{n}_r,\quad
\mathbf{y}_{d2}=\hat{h}_{sd}\mathbf{x}_{s21}+\tilde{h}_{sd}\mathbf{x}_{s21}+\mathbf{n}_{d2},
\text{ and}
\\
\mathbf{y}_d^r=\hat{h}_{sd}\mathbf{x}_{s22}+\hat{h}_{rd}\mathbf{x}_{r}+\tilde{h}_{sd}\mathbf{x}_{s22}+\tilde{h}_{rd}\mathbf{x}_{r}+\mathbf{n}_d^r.
\label{rdtx1x}
\end{gather}

\section{Achievable Rates} \label{sec:achievablerates}


In this section, we provide achievable rate expressions for AF and
DF relaying in both non-overlapped and overlapped transmission
scenarios described in Section \ref{sec:training}. Achievable rate
expressions are obtained by considering the estimate errors as
additional sources of Gaussian noise. Since Gaussian noise is the
worst uncorrelated additive noise for a Gaussian model
\cite{training}, \cite{Tong}, achievable rates given in this section
can be regarded as worst-case rates.

We first consider AF relaying scheme. The capacity of the AF relay
channel is the maximum mutual information between the transmitted
signal $\mathbf{x}_{s}$ and received signals $\mathbf{y}_d$ and
$\mathbf{y}_d^r$ given the estimates $\hat{h}_{sr}, \hat{h}_{sd},
\hat{h}_{rd}$:
\begin{equation} \label{eq:AFcap}
C=\sup_{p_{x_{s}}(\cdot )}\frac{1}{m}\emph{I}(\mathbf{x}_s
;\mathbf{y}_d,\mathbf{y}_d^r|\hat{h}_{sr},
 \hat{h}_{sd}, \hat{h}_{rd}).
\end{equation}
Note that this formulation presupposes that the destination has the
knowledge of $\hat{h}_{sr}$. Hence, we assume that the value of
$\hat{h}_{sr}$ is forwarded reliably from the relay to the
destination over low-rate control links.
In general, solving the optimization problem in (\ref{eq:AFcap}) and
obtaining the channel capacity is a difficult task. Therefore, we
concentrate on finding a lower bound on the capacity. A lower bound
is obtained by replacing the product of the estimate error and the
transmitted signal in the input-output relations with the worst-case
noise with the same correlation. In non-overlapped transmission, we
consider
\begin{equation} \label{eq:newnoisen}
\mathbf{z}_{d1}=\tilde{h}_{sd}\mathbf{x}_{s1}+\mathbf{n}_{d1}, \quad
\mathbf{z}_{r}=\tilde{h}_{sr}\mathbf{x}_{s2}+\mathbf{n}_r, \quad
\mathbf{z}_{d2}=\tilde{h}_{sd}\mathbf{x}_{s2}+\mathbf{n}_{d2},\quad
\text{and} \quad
\mathbf{z}_{d}^r=\tilde{h}_{rd}\mathbf{x}_{r}+\mathbf{n}_d^r,
\end{equation}
as the new noise vectors whose covariance matrices, respectively,
are
\begin{equation}\label{zrp}
E\{\mathbf{z}_{d1}\mathbf{z}_{d1}^\dagger\}=\sigma_{z_{d1}}^2\mathbf{I}=\sigma_{\tilde{h}_{sd}}^2E\{\mathbf{x}_{s1}\mathbf{x}_{s1}^\dagger\}+N_0\mathbf{I},
\quad
E\{\mathbf{z}_{r}\mathbf{z}_{r}^\dagger\}=\sigma_{z_{r}}^2\mathbf{I}=\sigma_{\tilde{h}_{sr}}^2E\{\mathbf{x}_{s2}\mathbf{x}_{s2}^\dagger\}+N_0\mathbf{I},
\end{equation}
\begin{equation}\label{zdp2}
E\{\mathbf{z}_{d2}\mathbf{z}_{d2}^\dagger\}=\sigma_{z_{d2}}^2\mathbf{I}=\sigma_{\tilde{h}_{sd}}^2E\{\mathbf{x}_{s2}\mathbf{x}_{s2}^\dagger\}+N_0\mathbf{I},
\quad
E\{\mathbf{z}_{d}^r{\mathbf{z}_{d}^r}^\dagger\}=\sigma_{z_{d}^r}^2\mathbf{I}=\sigma_{\tilde{h}_{rd}}^2E\{\mathbf{x}_{r}\mathbf{x}_{r}^\dagger\}+N_0\mathbf{I}.
\end{equation}
Similarly, in overlapped transmission, we define
\begin{equation}
\mathbf{z}_{d1}=\tilde{h}_{sd}\mathbf{x}_{s1}+\mathbf{n}_{d1}, \quad
\mathbf{z}_{r}=\tilde{h}_{sr}\mathbf{x}_{s21}+\mathbf{n}_r, \quad
\mathbf{z}_{d2}=\tilde{h}_{sd}\mathbf{x}_{s21}+\mathbf{n}_{d2},
\quad
\mathbf{z}_{d}^r=\tilde{h}_{sd}\mathbf{x}_{s22}+\tilde{h}_{rd}\mathbf{x}_{r}+\mathbf{n}_d^r,
\end{equation}
as noise vectors with covariance matrices
\begin{equation}\label{zrp1}
E\{\mathbf{z}_{d1}\mathbf{z}_{d1}^\dagger\}=\sigma_{z_{d1}}^2\mathbf{I}=\sigma_{\tilde{h}_{sd}}^2E\{\mathbf{x}_{s1}\mathbf{x}_{s1}^\dagger\}+N_0\mathbf{I},\quad
E\{\mathbf{z}_{r}\mathbf{z}_{r}^\dagger\}=\sigma_{z_{r}}^2\mathbf{I}=\sigma_{\tilde{h}_{sr}}^2E\{\mathbf{x}_{s21}\mathbf{x}_{s1}^\dagger\}+N_0\mathbf{I},
\end{equation}
\begin{equation}\label{zdp12}
E\{\mathbf{z}_{d2}\mathbf{z}_{d2}^\dagger\}=\sigma_{z_{d2}}^2\mathbf{I}=\sigma_{\tilde{h}_{sd}}^2E\{\mathbf{x}_{s21}\mathbf{x}_{s21}^\dagger\}+N_0\mathbf{I},\quad
E\{\mathbf{z}_{d}^r{\mathbf{z}_{d}^r}^\dagger\}=\sigma_{z_{d}^r}^2\mathbf{I}=\sigma_{\tilde{h}_{sd}}^2E\{\mathbf{x}_{s22}\mathbf{x}_{s22}^\dagger\}+\sigma_{\tilde{h}_{rd}}^2E\{\mathbf{x}_{r}\mathbf{x}_{r}^\dagger\}+N_0\mathbf{I}.
\end{equation}
An achievable rate expression is obtained by solving the following
optimization problem which requires finding the worst-case noise:
\begin{align}\label{ccc}
C\geqslant I_{AF}&=\inf_{p_{z_{d1}}(\cdot ),p_{z_{r}}(\cdot
),p_{z_{d2}}(\cdot ),p_{z_{d}^r}(\cdot )} \sup_{p_{x_{s}}(\cdot)}
\frac{1}{m}I(\mathbf{x}_s;\mathbf{y}_d,\mathbf{y}_d^r|\hat{h}_{sr},
 \hat{h}_{sd},\hat{h}_{rd}).
\end{align}
The following results provide $I_{AF}$ for both non-overlapped and
overlapped transmission scenarios.

\begin{prop:asympcap} \label{prop:asympcapaaa}
An achievable rate  of AF in non-overlapped transmission scheme is
given by
\begin{align}\label{AFC1}
I_{AF}=\frac{1}{m}E\left[(1-2\alpha)(m-2)\log\left(1+\frac{P_{s1}'|\hat{h}_{sd}|^2}{\sigma_{z_{d1}}^2}\right)
+\alpha(m-2)\log\left(1+\frac{P_{s1}'|\hat{h}_{sd}|^2}{\sigma_{z_{d2}}^2}+f\left[\frac{P_{s1}'|\hat{h}_{sr}|^2}{\sigma_{z_{r}}^2},\frac{P_{r}'|\hat{h}_{rd}|^2}{\sigma_{z_{d}^r}^2}\right]\right)\right]
\end{align}
where
\begin{equation}\label{ffunction}
f(x,y)=\frac{xy}{1+x+y}
\end{equation}
\begin{equation}\label{pssdzd1}
\frac{P_{s1}'|\hat{h}_{sd}|^2}{\sigma_{z_{d1}}^2} =
\frac{P_{s1}'|\hat{h}_{sd}|^2}{\sigma_{z_{d2}}^2}=\frac{\delta_s(1-\delta_s)m^2P_s^2\sigma_{sd}^4/(1-\alpha)}{(1-\delta_s)mP_s\sigma_{sd}^2N_0/(1-\alpha)+(m-2)(\sigma_{sd}^2\delta_smP_s+N_0)N_0}|w_{sd}|^2
\end{equation}
\begin{equation}\label{pssrzr1}
\frac{P_{s1}'|\hat{h}_{sr}|^2}{\sigma_{z_{r}}^2}=\frac{\delta_s(1-\delta_s)m^2P_s^2\sigma_{sr}^4/(1-\alpha)}{(1-\delta_s)mP_s\sigma_{sr}^2N_0/(1-\alpha)+(m-2)(\sigma_{sr}^2\delta_smP_s+N_0)N_0}|w_{sr}|^2
\end{equation}
\begin{equation}\label{prrdzdr1}
\frac{P_{r}'|\hat{h}_{rd}|^2}{\sigma_{z_{d}^r}^2}=\frac{\delta_r(1-\delta_r)m^2P_r^2\sigma_{rd}^4/\alpha}{(1-\delta_r)mP_r\sigma_{rd}^2N_0/\alpha+(m-2)(\sigma_{rd}^2\delta_rmP_r+N_0)N_0}|w_{rd}|^2.
\end{equation}
In the above equations and henceforth, $w_{sr}\sim\mathcal
{C}\mathcal {N}(0,1)$, $w_{sd}\sim\mathcal {C}\mathcal {N}(0,1)$,
$w_{rd}\sim\mathcal {C}\mathcal {N}(0,1)$ denote independent,
standard Gaussian random variables.
\end{prop:asympcap}

\emph{Proof}: Note that in non-overlapped AF relaying,
\begin{gather} \label{eq:mutualinfoAFn}
I(\mathbf{x}_s;\mathbf{y}_d,\mathbf{y}_d^r|\hat{h}_{sr},
 \hat{h}_{sd},\hat{h}_{rd}) = I(\mathbf{x}_{s1};\mathbf{y}_{d1}|\hat{h}_{sd}) +
I(\mathbf{x}_{s2};\mathbf{y}_{d2},\mathbf{y}_d^r|\hat{h}_{sr},
\hat{h}_{sd},\hat{h}_{rd})
\end{gather}
where the first mutual expression on the right-hand side of
(\ref{eq:mutualinfoAFn}) is for the direct transmission and the
second is for the cooperative transmission. In the direct
transmission, we have
\begin{gather}
\mathbf{y}_{d1}=\hat{h}_{sd}\mathbf{x}_{s1}+\z_{d1}.
\end{gather}
In this setting, it is well-known that the worst-case noise
$\z_{d1}$ is Gaussian \cite{training} and $\x_{s1}$ with independent
Gaussian components achieves
\begin{gather} \label{eq:directAFn}
\inf_{p_{z_{d1}}(\cdot )} \sup_{p_{x_{s1}}(\cdot)}
I(\mathbf{x}_{s1};\mathbf{y}_{d1}|\hat{h}_{sd}) =
E\left[(1-2\alpha)(m-2)\log\left(1+\frac{P_{s1}'|\hat{h}_{sd}|^2}{\sigma_{z_{d1}}^2}\right)\right].
\end{gather}
Therefore, we now concentrate on the cooperative phase. For better
illustration, we rewrite the channel input-output relationships in
(\ref{sdtx1}) and (\ref{rdtx1}) for each symbol:
\begin{equation}\label{yri}
y_r[i]=\hat{h}_{sr}x_{s2}[i]+z_{r}[i], \quad
y_{d2}[i]=\hat{h}_{sd}x_{s2}[i]+z_{d2}[i],
\end{equation}
for $i=1+(1-2\alpha)(m-2),...,(1-\alpha)(m-2)$, and
\begin{equation}\label{ydri}
y_d^r[i]=\hat{h}_{rd}x_{r}[i]+z_{d}^r[i],
\end{equation}
for $i=(1-\alpha)(m-2)+1,..., m-2$. In AF, the signals received and
transmitted by the relay have following relation:
\begin{equation}
x_{r}[i]=\beta y_r[i-\alpha(m-2)], \quad \text{where} \quad
\beta\leqslant\sqrt{\frac{E[|x_r|^2]}{|\hat{h}_{sr}|^2E[|x_{s2}|^2
]+E[|z_{r}|^2]}}.
\end{equation}
Now, we can write the channel in the vector form
\begin{eqnarray}\label{vec}
\underbrace{\left( \begin{array}{ccc}
y_{d2}[i] \\
y_d^r[i+\alpha(m-2)]  \\
\end{array} \right)}_{\mathbf{\check{y}}_d[i]}
=\underbrace{\left( \begin{array}{ccc}
\hat{h}_{sd} \\
\hat{h}_{rd}\beta\hat{h}_{sr} \\
\end{array} \right)}_{A}x_s[i]+
\underbrace{\left( \begin{array}{ccc}
0 &1&0 \\
\hat{h}_{rd}\beta&0&1 \\
\end{array} \right)}_{B}\underbrace{\left( \begin{array}{ccc}
z_{r}[i] \\
z_{d2}[i]\\
z_{d}^r[i+\alpha(m-2)]
\end{array} \right)}_{\mathbf{z}[i]}
\end{eqnarray}
for $i=1+(1-2\alpha)(m-2),...,(1-\alpha)(m-2)$,
With the  above notation, we can write the input-output mutual
information as
\begin{align}
\hspace{-.2cm}I(\mathbf{x}_{s2};\mathbf{y}_{d2},\mathbf{y}_d^r|\hat{h}_{sr},
 \hat{h}_{sd},\hat{h}_{rd})=\sum_{i =1+(1-2\alpha)(m-2)}^{(1-\alpha)(m-2)} \!\!\!\!\!\!I(x_s[i];\mathbf{\check{y}}_d[i]|\hat{h}_{sr},
 \hat{h}_{sd},\hat{h}_{rd})
 =\alpha(m-2) I(x_s;\mathbf{\check{y}}_d|\hat{h}_{sr},
 \hat{h}_{sd},\hat{h}_{rd}) \label{eq:simplifiedmutualinfo}
\end{align}
where in (\ref{eq:simplifiedmutualinfo}) we removed the dependence
on $i$ without loss of generality. Note that $\mathbf{\check{y}}$ is
defined in (\ref{vec}). Now, we can calculate the worst-case
capacity by proving that Gaussian distribution for $z_{r}$,
$z_{d2}$, and $z_{d}^r$ provides the worst case. We employ
techniques similar to that in~\cite{training}. Any set of particular
distributions for $z_{r}$, $z_{d2}$, and $z_{d}^r$ yields an upper
bound on the worst case. Let us choose $z_{r}$, $z_{d2}$, and
$z_{d}^r$ to be zero mean complex Gaussian distributed. Then as in
~\cite{jnl},
\begin{align}\label{cworst1}
\inf_{p_{z_{r}}(\cdot ),p_{z_{d2}}(\cdot ),p_{z_{d}^r}(\cdot )}
\sup_{p_{x_{s2}}(\cdot)} I(x_s;\mathbf{\check{y}}_d|\hat{h}_{sr},
 \hat{h}_{sd},\hat{h}_{rd}) \le  E \log
 \det\left(\mathbf{I}+(E(|x_s|^2)AA^\dagger)(BE[\z
 \z^\dagger]B^\dagger)^{-1}\right)
\end{align}
where the expectation is with respect to the fading estimates. To
obtain a lower bound, we compute the mutual information for the
channel in (\ref{vec}) assuming that $x_s$ is a zero-mean complex
Gaussian with variance $E(|x_s|^2)$, but the distributions of noise
components $z_{r}$, $z_{d2}$, and $z_{d}^r$ are arbitrary. In this
case, we have
\begin{align}\label{I1}
\emph{I}(x_s;\mathbf{\check{y}}_d;|\hat{h}_{sr},
\hat{h}_{sd},\hat{h}_{rd})
 &=\emph{h}(x_s|\hat{h}_{sr},
 \hat{h}_{sd},\hat{h}_{rd})-\emph{h}(x_s|\mathbf{\check{y}}_d,\hat{h}_{sr},
 \hat{h}_{sd},\hat{h}_{rd})\nonumber\\
& \geqslant \log\pi e E(|x_s|^2)-\log\pi e \,
var(x_s|\mathbf{\check{y}}_d,\hat{h}_{sr},
 \hat{h}_{sd},\hat{h}_{rd})
\end{align}
where the inequality is due to the fact that Gaussian distribution
provides the largest entropy and hence
$\emph{h}(x_s|\mathbf{\check{y}}_d,\hat{h}_{sr},
 \hat{h}_{sd},\hat{h}_{rd}) \le \log\pi e \,
var(x_s|\mathbf{\check{y}}_d,\hat{h}_{sr},
 \hat{h}_{sd},\hat{h}_{rd})$. From \cite{training}, we
know that
\begin{align}\label{cov}
var (x_s|\mathbf{\check{y}}_d,\hat{h}_{sr},
 \hat{h}_{sd},\hat{h}_{rd}) \leqslant
 E\left[(x_s-\hat{x}_{s})(x_s-\hat{x}_{s})^\dagger | \hat{h}_{sr},
 \hat{h}_{sd},\hat{h}_{rd}\right]
\end{align}
for any estimate $\hat{x}_{s}$ given
$\mathbf{\check{y}}_d,\hat{h}_{sr},
 \hat{h}_{sd}, \text{ and }\hat{h}_{rd}$. If we substitute the LMMSE estimate
$\hat{x}_{s}=R_{xy}R_{y}^{-1}\mathbf{\check{y}}_d$ into (\ref{I1})
and (\ref{cov}), we obtain \footnote{Here, we use the property that
$\det(\mathbf{I}+\mathbf{A}\mathbf{B})=\det(\mathbf{I}+\mathbf{B}\mathbf{A})$.}
\begin{align}\label{eq:mutualinfolowerbound}
I(x_s;\mathbf{\check{y}}_d|\hat{h}_{sr},\hat{h}_{sd},\hat{h}_{rd})
\!\ge \!E \log
 \det\left(\mathbf{I}+(E[|x_s|^2]AA^\dagger)(BE[\z
 \z^\dagger]B^\dagger)^{-1}\right).
\end{align}
Since the lower bound (\ref{eq:mutualinfolowerbound}) applies for
any noise distribution, we can easily see that
\begin{gather} \label{cworst2}
\inf_{p_{z_{r}}(\cdot ),p_{z_{d2}}(\cdot ),p_{z_{d}^r}(\cdot )}
\sup_{p_{x_{s2}}(\cdot)} I(x_s;\mathbf{\check{y}}_d|\hat{h}_{sr},
 \hat{h}_{sd},\hat{h}_{rd})\geqslant E \log
 \det\left(\mathbf{I}+(E[|x_s|^2]AA^\dagger)(BE[\z
 \z^\dagger]B^\dagger)^{-1}\right).
\end{gather}
From (\ref{cworst1}) and (\ref{cworst2}), we conclude that
\begin{align} \label{cworst}
\inf_{p_{z_{r}}(\cdot ),p_{z_{d2}}(\cdot ),p_{z_{d}^r}(\cdot )}
\sup_{p_{x_{s2}}(\cdot)} I(x_s;\mathbf{\check{y}}_d|\hat{h}_{sr},
 \hat{h}_{sd},\hat{h}_{rd}) &= E\log
 \det\left(\mathbf{I}+(E[|x_s|^2]AA^\dagger)(BE[\z
 \z^\dagger]B^\dagger)^{-1}\right)
 \\
&=E\left[\log\left(1+\frac{P_{s1}'|\hat{h}_{sd}|^2}{\sigma_{z_{d2}}^2}+
f\left[\frac{P_{s1}'|\hat{h}_{sr}|^2}{\sigma_{z_{r}}^2},\frac{P_{r}'|\hat{h}_{rd}|^2}{\sigma_{z_{d}^r}^2}\right]\right)\right].
\label{eq:cworstlog}
\end{align}
In (\ref{eq:cworstlog}), $P_{s1'}$ and $P_{r}'$ are the powers of
source and relay symbols and are given in (\ref{xsp2}) and
(\ref{xrp2}). Moreover, $\sigma_{z_{d2}}^2, \sigma_{z_{r}}^2,
\sigma_{z_{d}^r}^2$ are the variances of the noise components
defined in (\ref{eq:newnoisen}). Now, combining (\ref{ccc}),
(\ref{eq:mutualinfoAFn}), (\ref{eq:directAFn}), and
(\ref{eq:cworstlog}), we obtain the achievable rate expression in
(\ref{AFC1}). Note that (\ref{pssdzd1})--(\ref{prrdzdr1}) are
obtained by using the expressions for the channel estimates in
(\ref{est})--(\ref{eq:hrdest}) and noise variances in (\ref{zrp})
and (\ref{zdp2}). \hfill$ \square$
\begin{prop:asympcap} \label{prop:overlapped}
An achievable rate of AF in overlapped transmission scheme is given
by
\begin{align}\label{cafover1}
I_{AF} =\frac{1}{m}E
\Bigg[(1-2\alpha)(m-2)\log(1+\frac{P_{s2}'|\hat{h}_{sd}|^2}{\sigma_{z_{d1}}^2})+&(m-2)\alpha
\log\Bigg(1+\frac{P_{s2}'|\hat{h}_{sd}|^2}{\sigma_{z_{d2}}^2}+f\bigg(\frac{P_{s2}'|\hat{h}_{sr}|^2}{\sigma_{z_{r}}^2},\frac{P_{r}'|\hat{h}_{rd}|^2}{\sigma_{z_{d}^{r}}^2}\bigg)
\nonumber\\
&\!\!\!\!\!\!\!\!\!\!\!+q\bigg(\frac{P_{s2}'|\hat{h}_{sd}|^2}{\sigma_{z_{d2}}^2},\frac{P_{s2}'|\hat{h}_{sd}|^2}{\sigma_{z_{d}^{r}}^2},\frac{P_{s2}'|\hat{h}_{sr}|^2}{\sigma_{z_{r}}^2},\frac{P_{r}'|\hat{h}_{rd}|^2}{\sigma_{z_{d}^{r}}^2}\bigg)
\Bigg) \Bigg]
\end{align}
where  $q(.)$ is defined as $q(a,b,c,d)=\frac{(1+a)b(1+c)}{1+c+d}$.
Moreover
\begin{equation}\label{pssdzd2}
\frac{P_{s2}'|\hat{h}_{sd}|^2}{\sigma_{z_{d1}}^2}=\frac{P_{s2}'|\hat{h}_{sd}|^2}{\sigma_{z_{d2}}^2}=\frac{\delta_s(1-\delta_s)m^2P_s^2\sigma_{sd}^4}{(1-\delta_s)mP_s\sigma_{sd}^2N_0+(m-2)(\sigma_{sd}^2\delta_smP_s+N_0)N_0}|w_{sd}|^2
\end{equation}
\begin{equation}\label{pssrzr2}
\frac{P_{s2}'|\hat{h}_{sr}|^2}{\sigma_{z_{r}}^2}=\frac{\delta_s(1-\delta_s)m^2P_s^2\sigma_{sr}^4}{(1-\delta_s)mP_s\sigma_{sr}^2N_0+(m-2)(\sigma_{sr}^2\delta_smP_s+N_0)N_0}|w_{sr}|^2
\end{equation}
{\small{
\begin{eqnarray}\label{pssdzdr2}
\hspace{-1.3cm}\frac{P_{s2}'|\hat{h}_{sd}|^2}{\sigma_{z_{d}^{r}}^2}=\frac{\delta_s(1-\delta_s)m^2P_s^2\sigma_{sd}^4(\sigma_{rd}^2\delta_rmP_r+N_0)|w_{sd}|^2}{(m\!\!-\!\!2)
(\sigma_{sd}^2\delta_smP_s\!+\!N_0)(\sigma_{rd}^2\delta_rmP_r\!+\!N_0)N_0
+(1\!-\!\delta_r)mP_r\sigma_{rd}^2N_0(\sigma_{sd}^2\delta_smP_s+N_0)/
\alpha
+(1-\delta_s)mP_s\sigma_{sd}^2N_0(\sigma_{rd}^2\delta_rmP_r+N_0)}
\end{eqnarray}}}
{\small{
\begin{eqnarray}\label{prrdzdr2}
\hspace{-1.3cm}\frac{P_{r}'|\hat{h}_{rd}|^2}{\sigma_{z_{d}^{r}}^2}=\frac{\delta_r(1-\delta_r)m^2P_r^2\sigma_{rd}^4(\sigma_{sd}^2\delta_smP_s+N_0)/\alpha
|w_{rd}|^2}
{(m\!\!-\!\!2)(\sigma_{sd}^2\delta_smP_s\!+\!N_0)(\sigma_{rd}^2\delta_rmP_r\!+\!N_0)N_0
+(1-\delta_r)mP_r\sigma_{rd}^2N_0(\sigma_{sd}^2\delta_smP_s+N_0)/\alpha
+(1-\delta_s)mP_s\sigma_{sd}^2N_0(\sigma_{rd}^2\delta_rmP_r+N_0)}
\end{eqnarray}}}

\end{prop:asympcap}

\emph{Proof}: Note that the only difference between the overlapped
and non-overlapped transmissions is that source continues its
transmission as the relay transmits. As a result, the power of each
source symbol is now $P'_{s2}$ given in (\ref{xsp2}). Additionally,
when both the source and relay are transmitting, the received signal
at the destination is
$\mathbf{y}_d^r=\hat{h}_{sd}\mathbf{x}_{s22}+\hat{h}_{rd}\mathbf{x}_{r}+
\tilde{h}_{sd}\mathbf{x}_{s22}+\tilde{h}_{rd}\mathbf{x}_{r}+\mathbf{n}_d^r.$
The input-output mutual information in one block is 
\begin{gather} \label{eq:mutualinfoAFo}
I(\mathbf{x}_s;\mathbf{y}_d,\mathbf{y}_d^r|\hat{h}_{sr},
 \hat{h}_{sd},\hat{h}_{rd}) = I(\mathbf{x}_{s1};\mathbf{y}_{d1}|\hat{h}_{sd}) +
I(\mathbf{x}_{s21},\x_{s22};\mathbf{y}_{d2},\mathbf{y}_d^r|\hat{h}_{sr},
\hat{h}_{sd},\hat{h}_{rd}).
\end{gather}

The first term on the right-hand-side of (\ref{eq:mutualinfoAFo})
corresponds to the mutual information of the direct transmission and
is the same as that in non-overlapped transmission. Hence, the
worst-case rate expression obtained in the proof of Theorem
\ref{prop:asympcapaaa} is valid for this case as well. In the
cooperative phase, the input-output relation for each symbol can be
written in the following matrix form: {\small
\begin{eqnarray}\label{vecover}
\underbrace{\left( \begin{array}{ccc}
y_{d2}[i] \\
y_d^r[i+\alpha(m-2)]  \\
\end{array} \right)}_{\mathbf{\check{y}}_d[i]}
=\underbrace{\left( \begin{array}{ccc}
\hat{h}_{sd}&0 \\
\hat{h}_{rd}\beta\hat{h}_{sr} & \hat{h}_{sd} \\
\end{array} \right)}_{A}\underbrace{\left(\begin{array}{ccc}
x_s[i]\\
x_s[i+\alpha(m-2)]\\
\end{array}\right)}_{\mathbf{\check{x}}_s[i]}
+\underbrace{\left( \begin{array}{ccc}
0 &1&0 \\
\hat{h}_{rd}\beta&0&1 \\
\end{array} \right)}_{B}\underbrace{\left( \begin{array}{ccc}
z_{r}[i] \\
z_{d2}[i]\\
z_{d}^r[i+\alpha(m-2)]
\end{array} \right)}_{\mathbf{z}[i]}
\end{eqnarray}}
where $i=1+(1-2\alpha)(m-2),..., (1-\alpha)(m-2)$ and $
\beta\leqslant\sqrt{\frac{E[|x_r|^2]}{|\hat{h}_{sr}|^2E[|x_s|^2
]+E[|z_{r}|^2]}}.$ Note that we have defined
$\mathbf{x}_{s}=[\mathbf{x}_{s1}^T,\mathbf{x}_{s21}^T,\mathbf{x}_{s22}^T]^T$,
and the expression in (\ref{vecover}) uses the property that
$x_{21}(j) = x_s(j + (1-2\alpha)(m-2))$ and $x_{s22}(j) = x_{s}(j +
(1-\alpha)(m-2))$ for $j = 1,\ldots, \alpha(m-2)$. The input-output
mutual information in the cooperative phase can now be expressed as
\begin{align}
\hspace{-1cm}I(\mathbf{x}_{s21},
\x_{s22};\mathbf{y}_{d2},\mathbf{y}_d^r|\hat{h}_{sr},
 \hat{h}_{sd},\hat{h}_{rd})=\!\!\!\!\!\sum_{i = 1+(1-2\alpha)(m-2)}^{(1-\alpha)(m-2)} \!\!\!\!\!\!I(\mathbf{\check{x}}_s[i];\mathbf{\check{y}}_d[i]|\hat{h}_{sr},
 \hat{h}_{sd},\hat{h}_{rd})
=\alpha(m-2)
I(\mathbf{\check{x}_s};\mathbf{\check{y}}_d|\hat{h}_{sr}
 \hat{h}_{sd},\hat{h}_{rd}) \label{eq:simplifiedmutualinfo1}
\end{align}
where in (\ref{eq:simplifiedmutualinfo1}) we removed the dependence
on $i$ without loss of generality. Note that $\mathbf{\check{x}}$
and $\mathbf{\check{y}}$ are defined in (\ref{vecover}). As shown in
proof of Theorem \ref{prop:asympcapaaa}, the worst-case achievable
rate for cooperative transmission is
\begin{gather} \label{cworstv}
\inf_{p_{z_{r}}(\cdot ),p_{z_{d2}}(\cdot ),p_{z_{d}^r}(\cdot )}
\sup_{p_{x_{s2}}(\cdot)}
I(\mathbf{\check{x}_s};\mathbf{\check{y}}_d|\hat{h}_{sr},
 \hat{h}_{sd},\hat{h}_{rd}) =
E\log
 \det\left(\mathbf{I}+(E[\mathbf{\check{x}_s}\mathbf{\check{x}_s}^\dagger]AA^\dagger)(BE[\z
 \z^\dagger]B^\dagger)^{-1}\right).
\end{gather}
Using the definitions in (\ref{vecover}) and evaluating the
$\log\det$ expression in (\ref{cworstv}), and combining the direct
transmission worst-case achievable rate, we arrive to
(\ref{cafover1}). (\ref{pssdzd2})--(\ref{prrdzdr2}) are obtained by
using the expressions for the channel estimates in
(\ref{est})--(\ref{eq:hrdest}) and noise variances in (\ref{zrp1})
and (\ref{zdp12}).
\hfill $\square$

Next, we consider DF relaying scheme. In DF, there are two different
coding approaches \cite{jnlbook}, namely repetition coding and
parallel channel coding. We first consider repetition channel coding
scheme. The following results provide achievable rate expressions in
both non-overlapped and overlapped transmission scenarios.
\begin{prop:asympcap} \label{prop:DFRnonover}
An achievable rate expression for DF with repetition channel coding
for non-overlapped transmission scheme is given by
\begin{equation}\label{DFC1}
I_{DFr}=
\frac{(1-2\alpha)(m-2)}{m}E\left[\log\left(1+\frac{P_{s1}'|\hat{h}_{sd}|^2}{\sigma_{z_{d1}}^2}\right)\right]
+\frac{\alpha(m-2)}{m} \min\{I_1,I_2\}
\end{equation}
where
\begin{equation}\label{C1}
I_1=E\bigg[\log\Big(1+\frac{P_{s1}'|\hat{h}_{sr}|^2}{\sigma_{z_{r}}^2}\Big)\bigg],
\quad \text{and} \quad
%
I_2=E\bigg[\log\Big(1+\frac{P_{s1}'|\hat{h}_{sd}|^2}{\sigma_{z_{d2}}^2}+\frac{P_{r}'|\hat{h}_{rd}|^2}{\sigma_{z_{d}^r}^2}\Big)\bigg].
\end{equation}
Moreover,
$\frac{P_{s1}'|\hat{h}_{sd}|^2}{\sigma_{z_{d1}}^2}$,$\frac{P_{s1}'|\hat{h}_{sd}|^2}{\sigma_{z_{d2}}^2}$,
$\frac{P_{s1}'|\hat{h}_{sr}|^2}{\sigma_{z_{r}}^2},$ and
 $\frac{P_{r}'|\hat{h}_{rd}|^2}{\sigma_{z_{d}^r}^2}$ are the same as defined in
(\ref{pssdzd1})--(\ref{prrdzdr1}).
\end{prop:asympcap}

\emph{Proof}: For DF with repetition coding in non-overlapped
transmission, an achievable rate expression is
\begin{align}\label{IDF1}
I(\mathbf{x}_{s1};\mathbf{y}_{d1}|\hat{h}_{sd}) +
\min\left\{\emph{I}(\mathbf{x}_{s2};\mathbf{y}_r|\hat{h}_{sr}),\emph{I}(\mathbf{x}_{s2};\mathbf{y}_d,\mathbf{y}_d^r|\hat{h}_{sd},\hat{h}_{rd})\right\}.
\end{align}
Note that the first and second mutual information expressions in
(\ref{IDF1}) are for the direct transmission between the source and
destination, and direct transmission between the source and relay,
respectively. Therefore, as in the proof of Theorem
\ref{prop:asympcapaaa}, the worst-case achievable rates can be
immediately seen to be equal to the first term on the right-hand
side of (\ref{DFC1}) and $I_1$, respectively.

In repetition coding, after successfully decoding the source
information, the relay transmits the same codeword as the source. As
a result, the input-output relation in the cooperative phase can be
expressed as 
\begin{eqnarray}\label{vec1}
\underbrace{\left( \begin{array}{ccc}
y_d[i] \\
y_d^r[i+\alpha(m-2)]  \\
\end{array} \right)}_{\mathbf{\check{y}}_d[i]}
=\underbrace{\left( \begin{array}{ccc}
\hat{h}_{sd} \\
\hat{h}_{rd}\beta \\
\end{array} \right)}_{A}x_s[i]+
\underbrace{\left(
\begin{array}{ccc}
z_{d2}[i]\\
z_{d}^r[i+\alpha(m-2)]
\end{array} \right)}_{z[i]}.
\end{eqnarray}
where $ \beta\leq \sqrt{\frac{E[|x_r|^2]}{E[|x_s|^2]}}. $ From
(\ref{vec1}), it is clear that the knowledge of $\he_{sr}$ is not
required at the destination. We can easily see that (\ref{vec1}) is
a simpler expression than (\ref{vec}) in the AF case, therefore we
can adopt the same methods as employed in the proof of Theorem
\ref{prop:asympcapaaa} to show that
Gaussian noise is the worst noise and $I_2$ is the worst-case rate. 
\hfill $\square$

\begin{prop:asympcap} \label{prop:DFRover}
An achievable rate expression for DF with repetition channel coding
for overlapped transmission scheme is given by
\begin{equation}\label{DFC12}
I_{DFr}= \frac{(1-2\alpha)(m-2)}{m}E\left[
\log\left(1+\frac{P_{s2}'|\hat{h}_{sd}|^2}{\sigma_{z_{d1}}^2}\right)\right]
+\frac{(m-2)\alpha}{m} \min\{I_1,I_2\}
\end{equation}
where
\begin{align}\label{DFC11}
I_1=E\Bigg[\log\Bigg(1+\frac{P_{s2}'|\hat{h}_{sr}|^2}{\sigma_{z_{r}}^2}\Bigg)\Bigg],
I_2=E\Bigg[\log
\left(1+\frac{P_{s2}'|\hat{h}_{sd}|^2}{\sigma_{z_{d2}}^2}+\frac{P_{r}'|\hat{h}_{rd}|^2}{\sigma_{z_{d}^r}^2}
+\frac{P_{s2}'|\hat{h}_{sd}|^2}{\sigma_{z_{d}^r}^2}+
\frac{P_{s2}'|\hat{h}_{sd}|^2}{\sigma_{z_{d2}}^2}\frac{P_{s2}'|\hat{h}_{sd}|^2}{\sigma_{z_{d}^r}^2}\right)\Bigg].
\end{align}
$\frac{P_{s2}'|\hat{h}_{sd}|^2}{\sigma_{z_{d1}}^2}$,$\frac{P_{s2}'|\hat{h}_{sd}|^2}{\sigma_{z_{d2}}^2}$,
$\frac{P_{s2}'|\hat{h}_{sr}|^2}{\sigma_{z_{r}}^2}$,$\frac{P_{s2}'|\hat{h}_{sd}|^2}{\sigma_{z_{d}^r}^2}$,
$\frac{P_{r}'|\hat{h}_{rd}|^2}{\sigma_{z_{d}^r}^2}$ have the same
expressions as in (\ref{pssdzd2})--(\ref{prrdzdr2}).
\end{prop:asympcap}

\emph{Proof}: 
Note that in overlapped transmission, source transmits over the
entire duration of $(m-2)$ symbols, and hence the channel
input-output relation in the cooperative phase is expressed as
follows:
\begin{eqnarray}\label{vec1o}
\underbrace{\left( \begin{array}{ccc}
y_d[i] \\
y_d^r[i+\alpha(m-2)]  \\
\end{array} \right)}_{\check{\mathbf{y}}_d[i]}
=\underbrace{\left( \begin{array}{ccc}
\hat{h}_{sd} &0\\
\hat{h}_{rd}\beta & \hat{h}_{sd} \\
\end{array} \right)}_{A}\underbrace{\left(\begin{array}{ccc}
x_s[i]\\
x_s[i+\alpha(m-2)]\\
\end{array}\right)}_{\mathbf{\check{x}}_s[i]}+
\underbrace{\left(
\begin{array}{ccc}
z_{d22}[i]\\
z_{d2}^r[i+\alpha(m-2)]
\end{array} \right)}_{z[i]}.
\end{eqnarray}
The result is immediately obtained using the same techniques as in
the proof of Theorem \ref{prop:overlapped}. \hfill $\square$

Finally, we consider DF with parallel channel coding and assume that
non-overlapped transmission scheme is adopted. From \cite{Liang}, we
note that an achievable rate expression is
\begin{align*}
\min \{ &(1-\alpha)
\emph{I}(\mathbf{x}_s ;\mathbf{y}_r|\hat{h}_{sr}),(1-\alpha)
\emph{I}(\mathbf{x}_s ;\mathbf{y}_d|\hat{h}_{sd})+\alpha
\emph{I}(\mathbf{x}_r ;\mathbf{y}_d^r|\hat{h}_{rd})\}.
\end{align*}
Note that we do not have separate direct transmission in this
relaying scheme. Using similar methods as before, we obtain the
following result. The proof is omitted to avoid repetition.
\begin{prop:asympcap} \label{prop:asympcap}
An achievable rate of non-overlapped  DF with parallel channel
coding scheme is given by
\begin{align}\label{DFP}
I_{DFp}=\min \Bigg \{\frac{(1-\alpha)(m-2)}{m}E \left[
\log\left(1+\frac{P_{s1}'|\hat{h}_{sr}|^2}{\sigma_{z_{r}}^2}\right)\right],&
\frac{(1-\alpha)(m-2)}{m}E \left[\log\left(1+\frac{P_{s1}'|\hat{h}_{sd}|^2}{\sigma_{z_{d2}}^2}\right)\right]+\nonumber \\
&\frac{\alpha (m-2)}{m}E\left[\log
\left(1+\frac{P_{r}'|\hat{h}_{rd}|^2}{\sigma_{z_{d}^r}^2}\right)
\right]\Bigg\}
\end{align}
where  $\frac{P_{s1}'|\hat{h}_{sd}|^2}{\sigma_{z_{d2}}^2}$,
$\frac{P_{s1}'|\hat{h}_{sr}|^2}{\sigma_{z_{r}}^2},$ and
 $\frac{P_{r}'|\hat{h}_{rd}|^2}{\sigma_{z_{d}^r}^2}$ are given in
 (\ref{pssdzd1})-(\ref{prrdzdr1}). \hfill $\square$
\end{prop:asympcap}

\section{Optimal Resource Allocation} \label{sec:allocation}

Having obtained achievable rate expressions in Section
\ref{sec:achievablerates}, we now identify optimal resource
allocation strategies that maximize the rates. We consider three
resource allocation problems: 1) power allocation between the
training and data symbols; 2) time/bandwidth allocation to the
relay; 3) power allocation between the source and relay under a
total power constraint.

\begin{figure}
\begin{center}
\includegraphics[width = 0.6\textwidth]{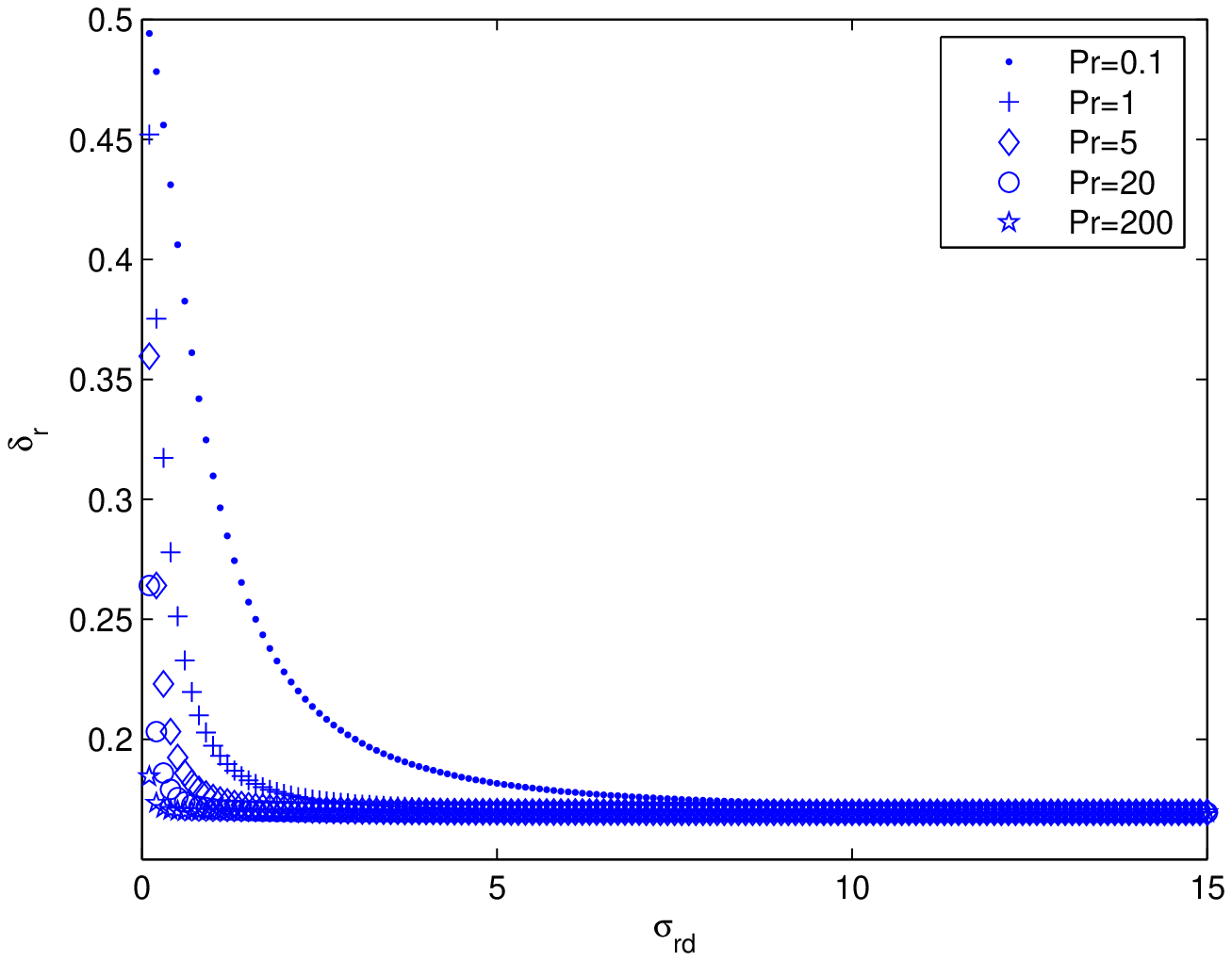}
\caption{$\delta_r$ vs. $\sigma_{rd}$ for different values of $P_r$
when $m =50$.} \label{fig:1}
\end{center}
\end{figure}

We first study how much power should be allocated for channel
training. 
In nonoverlapped AF, it can be seen that $\delta_r$ appears only in
$\frac{P_{r}'|\hat{h}_{rd}|^2}{\sigma_{z_{d}^r}^2}$ in the
achievable rate expression (\ref{AFC1}). Since $
f(x,y)=\frac{xy}{1+x+y} $ is a monotonically increasing function of
$y$ for fixed $x$, (\ref{AFC1}) is maximized by maximizing
$\frac{P_{r}'|\hat{h}_{rd}|^2}{\sigma_{z_{d}^r}^2}$. We can maximize
$\frac{P_{r}'|\hat{h}_{rd}|^2}{\sigma_{z_{d}^r}^2}$ by maximizing
the coefficient of the random variable $|w_{rd}|^2$ in
(\ref{prrdzdr1}), and the optimal $\delta_r$ is given below:

{\footnotesize
\begin{equation}\label{taor}
\delta_r^{opt}=\frac{-mP_r\sigma_{rd}^2-\alpha mN_0+2\alpha
N_0+\sqrt{\alpha(m-2)(m^2P_r\sigma_{rd}^2\alpha
N_0+m^2P_r^2\sigma_{rd}^4+\alpha m
N_0^2+mP_r\sigma_{rd}^2N_0-2mP_r\sigma_{rd}^2\alpha
N_0-2N_0\alpha)}}{mP_r\sigma_{rd}^2(-1+\alpha m-2\alpha)}.
\end{equation}
} Optimizing $\delta_s$ is more complicated as it is related to all
the terms in (\ref{AFC1}), and hence obtaining an analytical
solution is unlikely. A suboptimal solution is to maximize
$\frac{P_{s1}'|\hat{h}_{sd}|^2}{\sigma_{z_{d1}}^2}$ and
$\frac{P_{s1}'|\hat{h}_{sr}|^2}{\sigma_{z_{r}}^2}$ separately, and
obtain two solutions $\delta_{s,1}^{subopt}$ and
$\delta_{s,2}^{subopt}$, respectively. Note that expressions for
$\delta_{s,1}^{subopt}$ and $\delta_{s,2}^{subopt}$ are exactly the
same as that in (\ref{taor}) with $P_r$, $\alpha$, and $\sigma_{rd}$
replaced by $P_s$, ($1-\alpha$), and $\sigma_{sd}$ and
$\sigma_{sr}$, respectively.
When the source-relay channel is better than the source-destination
channel and the fraction of time over which direct transmission is
performed is small,
$\frac{P_{s1}'|\hat{h}_{sr}|^2}{\sigma_{z_{r}}^2}$ is a more
dominant factor and $\delta_{s,2}^{subopt}$ is a good choice for
training power allocation. Otherwise, $\delta_{s,1}^{subopt}$ might
be preferred. Note that in non-overlapped DF with repetition and
parallel coding, $\frac{P_{r}'|\hat{h}_{rd}|^2}{\sigma_{z_{d}^r}^2}$
is the only term that includes $\delta_r$. Therefore, similar
results and discussions apply. For instance, the optimal $\delta_r$
has the same expression as that in (\ref{taor}).
Figure \ref{fig:1} plots the optimal $\delta_r$ as a function of
$\sigma_{rd}$ for different relay power constraints $P_r$ when $m =
50$ and $\alpha=0.5$. It is observed in all cases that the allocated
training power monotonically decreases with improving channel
quality and converges to $\frac{\sqrt{\alpha(m-2)}-1}{\alpha
m-2\alpha-1} \approx 0.169$ which is independent of $P_r$.

\begin{figure}
\begin{center}
\includegraphics[width = 0.55\textwidth]{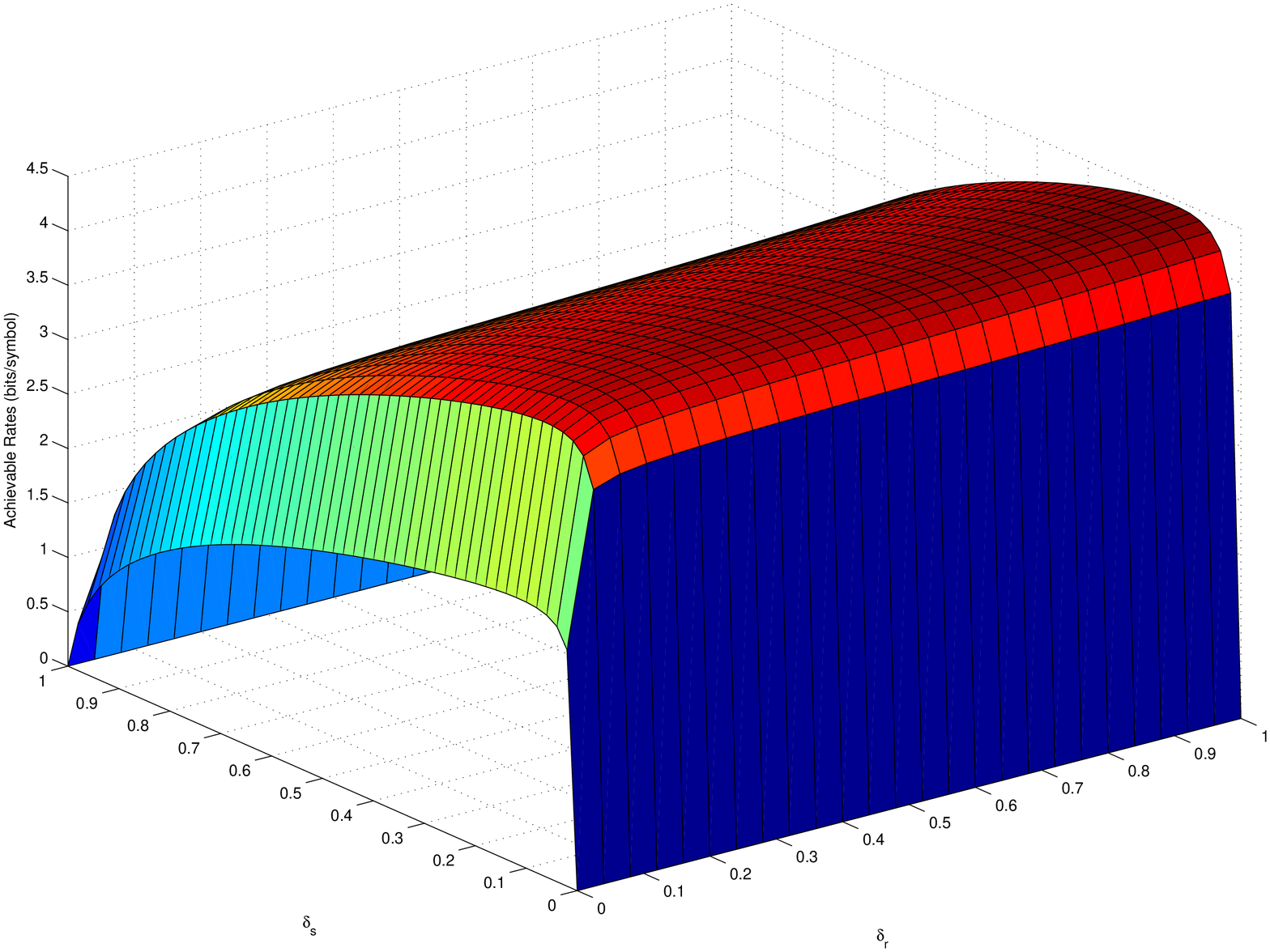}
\caption{Overlapped AF achievable rates vs. $\delta_s$ and
$\delta_r$ when $P_s=P_r=50$} \label{fig:ovtraf}
\end{center}
\end{figure}

\begin{figure}
\begin{center}
\includegraphics[width = \figsize\textwidth]{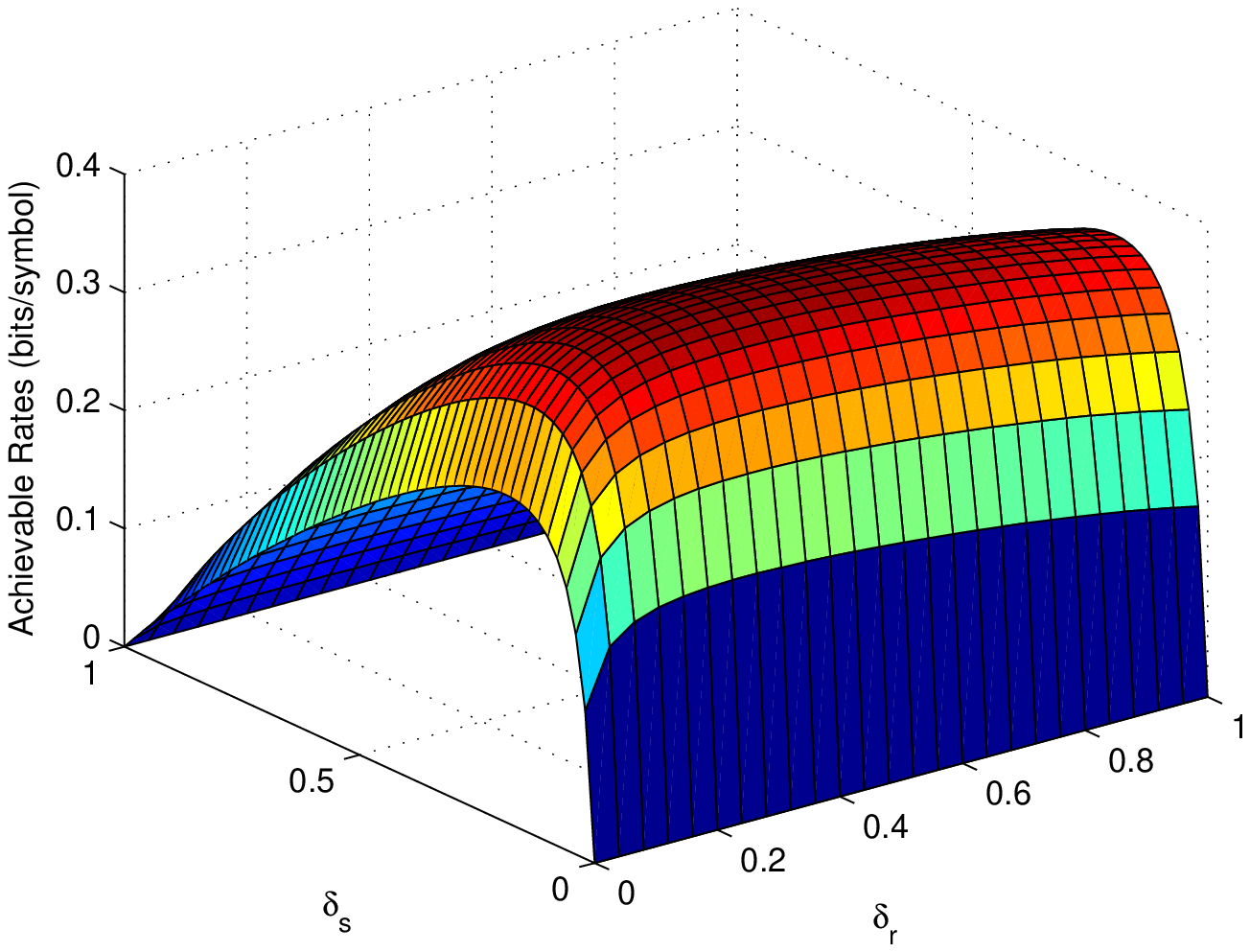}
\caption{Overlapped AF achievable rates vs. $\delta_s$ and
$\delta_r$ when $P_s=P_r=0.5$} \label{fig:ovtrafl}
\end{center}
\end{figure}

In overlapped transmission schemes, both $\delta_s$ and $\delta_r$
appear in more than one term in the achievable rate expressions.
Therefore, we resort to numerical results to identify the optimal
values. Figures \ref{fig:ovtraf} and \ref{fig:ovtrafl} plot the
achievable rates as a function of $\delta_s$ and $\delta_r$ for
overlapped AF. In both figures, we have assumed that $\sigma_{sd}=1,
\sigma_{sr}=2, \sigma_{rd}=1$ and $m=50, N_0=1,\alpha=0.5$. While
Fig. \ref{fig:ovtraf}, where $P_s=50$ and $P_r=50$, considers high
$\tsnr$s, we assume that $P_s=0.5$ and $P_r=0.5$ in Fig.
\ref{fig:ovtrafl}. In Fig. \ref{fig:ovtraf}, we observe that
increasing $\delta_s$ will increase achievable rate until $\delta_s
\approx 0.1$. Further increase in $\delta_s$ decreases the
achievable rates. On the other hand, rates always increase with
increasing $\delta_r$. This indicates that cooperation is not
beneficial in terms of achievable rates and direct transmission
should be preferred.
On the other hand, in the low-power regime considered in Fig.
\ref{fig:ovtrafl}, the optimal values of $\delta_s$ and $\delta_r$
are approximately 0.18 and 0.32, respectively. Hence, the relay in
this
case helps to improve the rates. 

\begin{figure}
\begin{center}
\includegraphics[width = \figsize\textwidth]{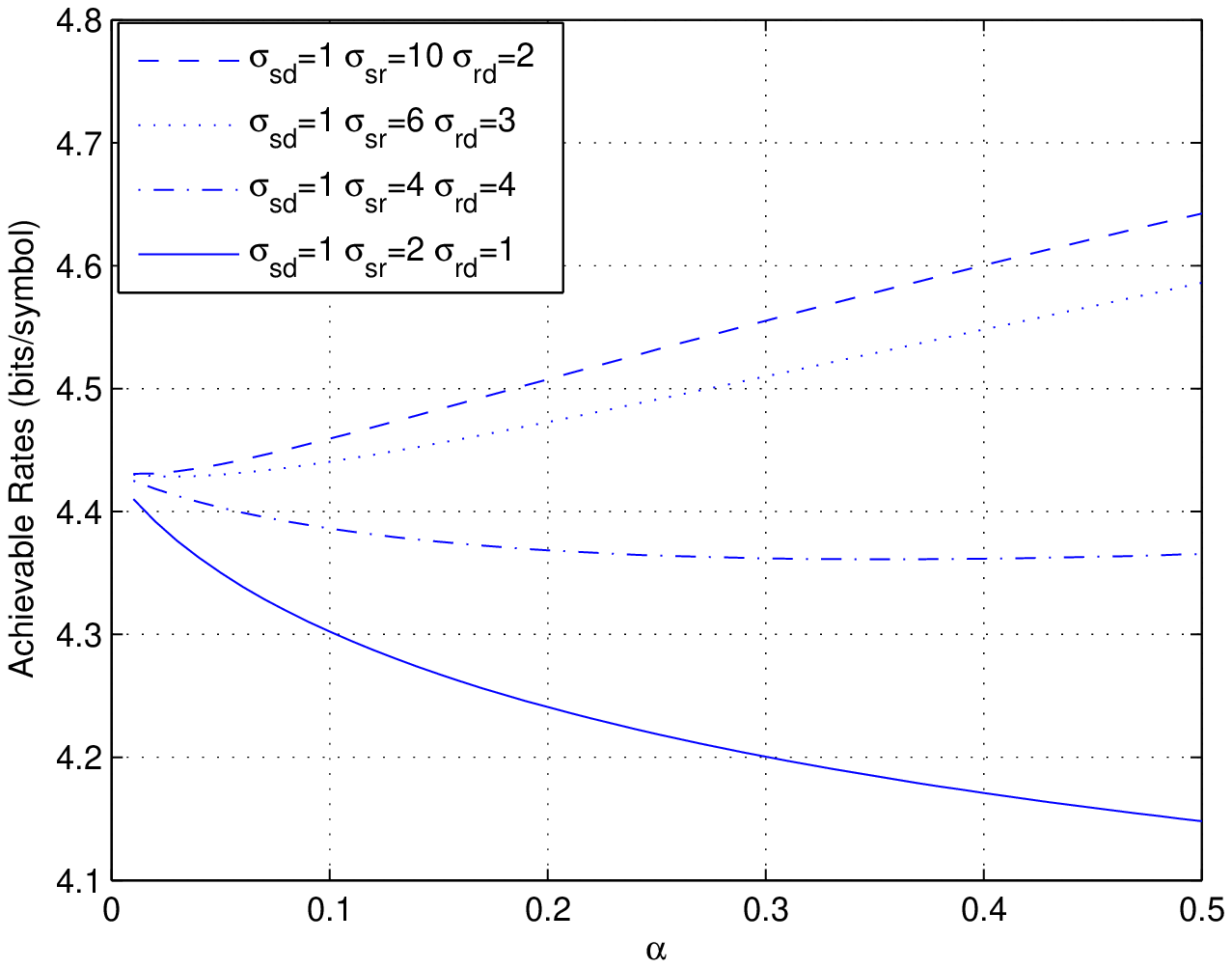}
\caption{ Overlapped AF achievable rate  vs. $\alpha$ when
$P_s=P_r=50, \delta_s=\delta_r=0.1$, $m=50$.} \label{fig:overAF1}
\end{center}
\end{figure}

\begin{figure}
\begin{center}
\includegraphics[width = \figsize\textwidth]{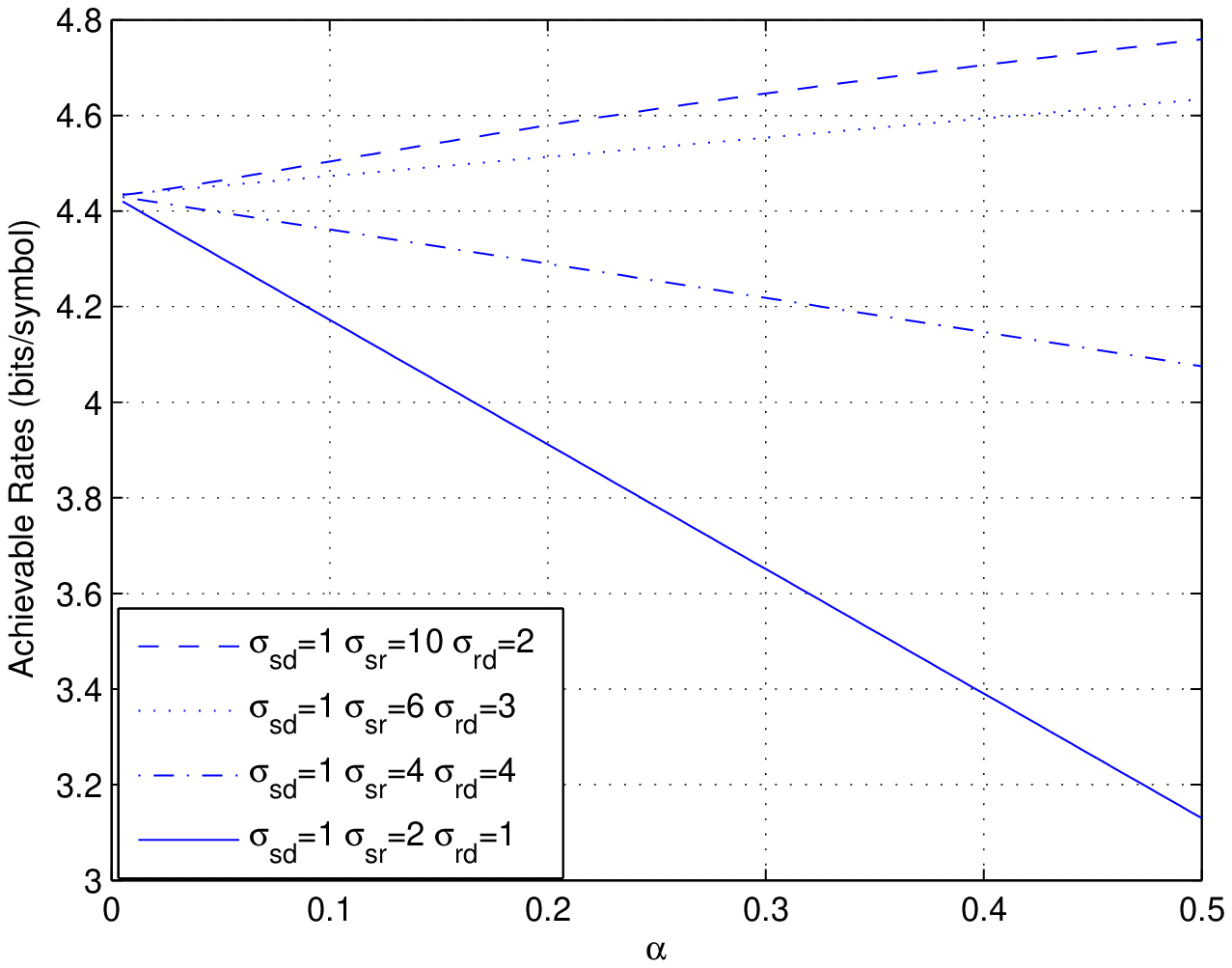}
\caption{ Overlapped DF with repetition coding achievable rate  vs.
$\alpha$ when $P_s=P_r=50, \delta_s=\delta_r=0.1$, $m=50$.}
\label{fig:overDF}
\end{center}
\end{figure}

\begin{figure}
\begin{center}
\includegraphics[width = \figsize\textwidth]{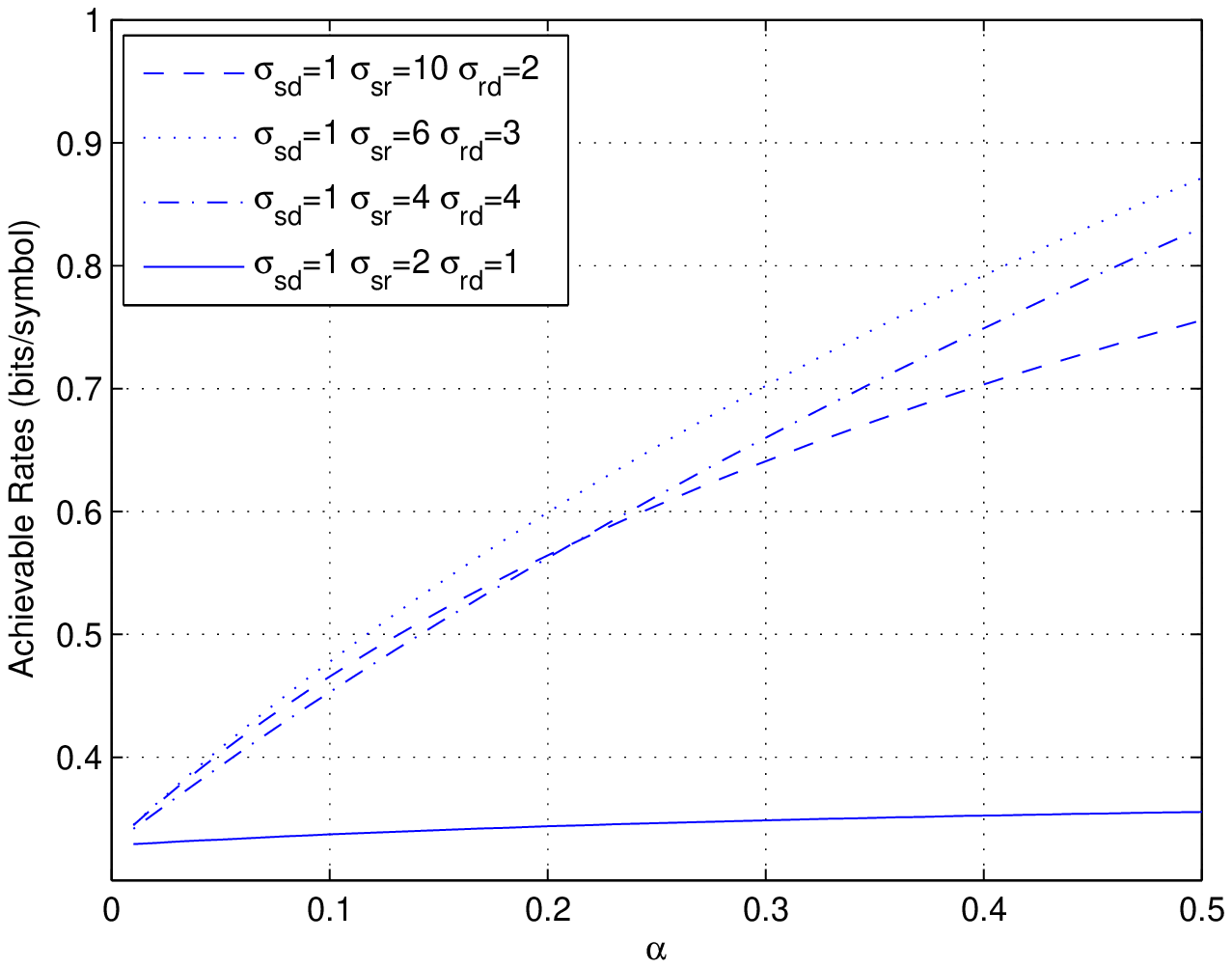}
\caption{ Overlapped AF achievable rate  vs. $\alpha$ when
$P_s=P_r=0.5, \delta_s=\delta_r=0.1$, $m=50$.} \label{fig:overAF2}
\end{center}
\end{figure}

\begin{figure}
\begin{center}
\includegraphics[width = \figsize\textwidth]{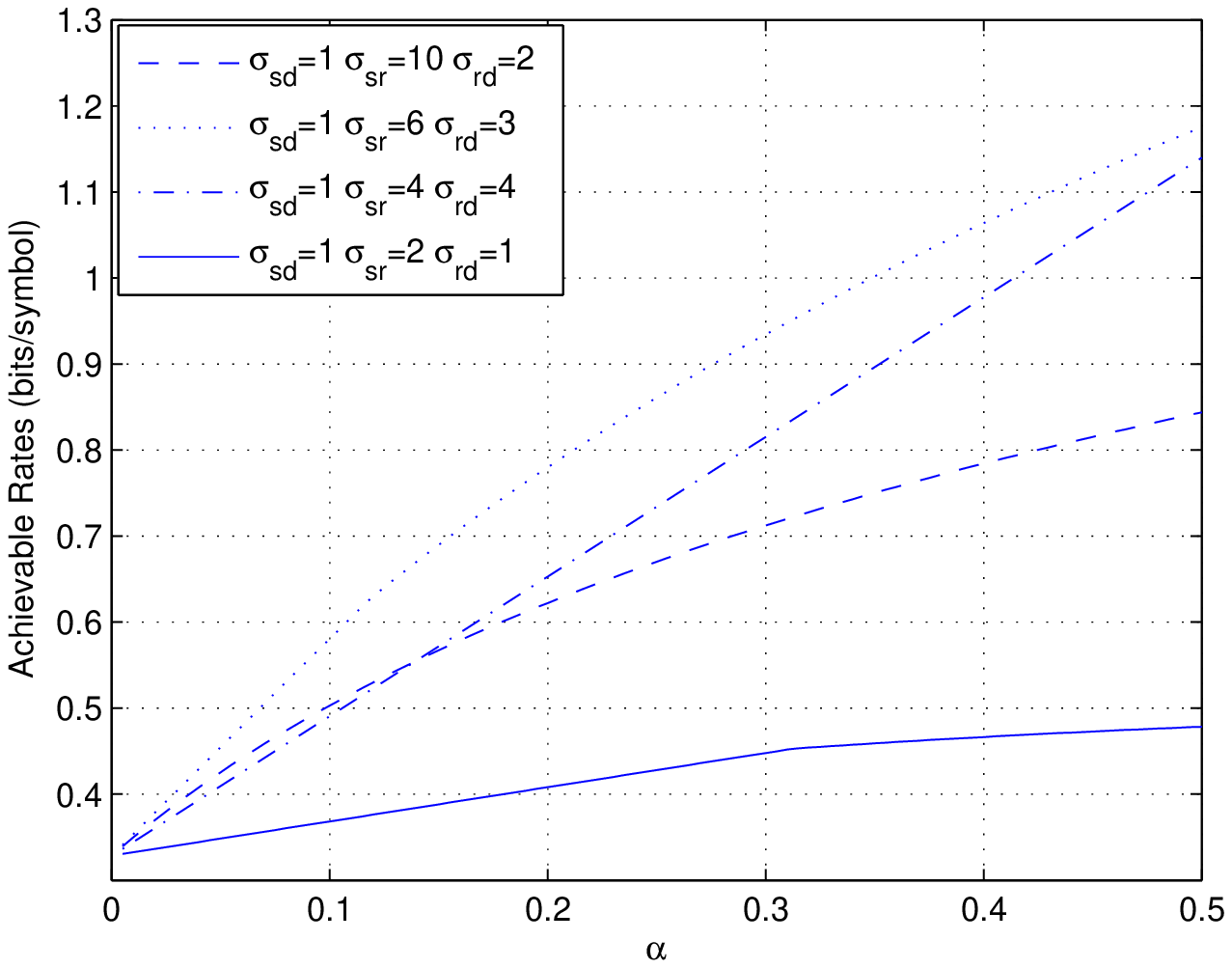}
\caption{ Overlapped DF with repetition coding achievable rate  vs.
$\alpha$ when $P_s=P_r=0.5, \delta_s=\delta_r=0.1$, $m=50$.}
\label{fig:overDF2}
\end{center}
\end{figure}

Next, we analyze the effect of the degree of cooperation on the
performance in AF and repetition DF. Figures
\ref{fig:overAF1}-\ref{fig:overDF2} plot the achievable rates as a
function of $\alpha$ which gives the fraction of total
time/bandwidth allocated to the relay. Achievable rates are obtained
for different channel qualities given by the standard deviations
$\sigma_{sd}, \sigma_{sr},$ and $\sigma_{rd}$ of the fading
coefficients. We observe that if the input power is high, $\alpha$
should be either $0.5$ or close to zero depending on the channel
qualities. On the other hand, $\alpha=0.5$ always gives us the best
performance at low $\tsnr$ levels regardless of the channel
qualities. Hence, while cooperation is beneficial in the low-$\tsnr$
regime, noncooperative transmissions might be optimal at high
$\tsnr$s. We note from Fig. \ref{fig:overAF1} that cooperation
starts being useful as the source-relay channel variance
$\sigma^2_{sr}$ increases. Similar results are also observed in Fig
\ref{fig:overDF}. Hence, the source-relay channel quality is one of
the key factors in determining the usefulness of cooperation in the
high $\tsnr$ regime.

\begin{figure}
\begin{center}
\includegraphics[width = \figsize\textwidth]{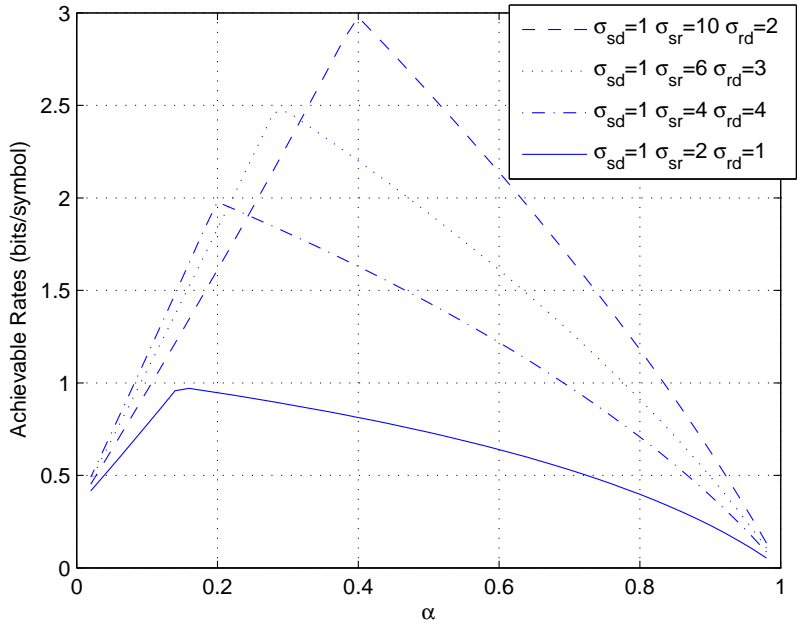}
\caption{ Non-overlapped DF parallel coding achievable rate  vs.
$\alpha$ when $P_s=P_r=0.5, \delta_s=\delta_r=0.1$, $m=50$.}
\label{fig:overDFP}
\end{center}
\end{figure}

In Fig. \ref{fig:overDFP}, we plot the achievable rates of DF
parallel channel coding, derived in Theorem \ref{prop:asympcap}. We
can see from the figure that the best performance is obtained when
the source-relay channel quality is high (i.e., when $\sigma_{sd} =
1, \sigma_{sr} = 10, \sigma_{rd} = 2$). Additionally, we observe
that as the source-relay channel improves, more resources need to be
allocated to the relay to achieve the best performance. We note that
significant improvements with respect to direct transmission (i.e.,
the case in which $\alpha \to 0$) are obtained. Finally, we can see
that when compared to AF and DF with repetition coding, DF with
parallel channel coding achieves higher rates. On the other hand, AF
and repetition coding DF have advantages in the implementation.
Obviously, the relay, which amplifies and forwards, has a simpler
task than that which decodes and forwards. Moreover, as pointed out
in \cite{tutorial}, if AF or repetition coding DF is employed in the
system, the architecture of the destination node is simplified
because the data arriving from the source and relay can be combined
rather than stored separately.

\begin{figure}
\begin{center}
\includegraphics[width = \figsize\textwidth]{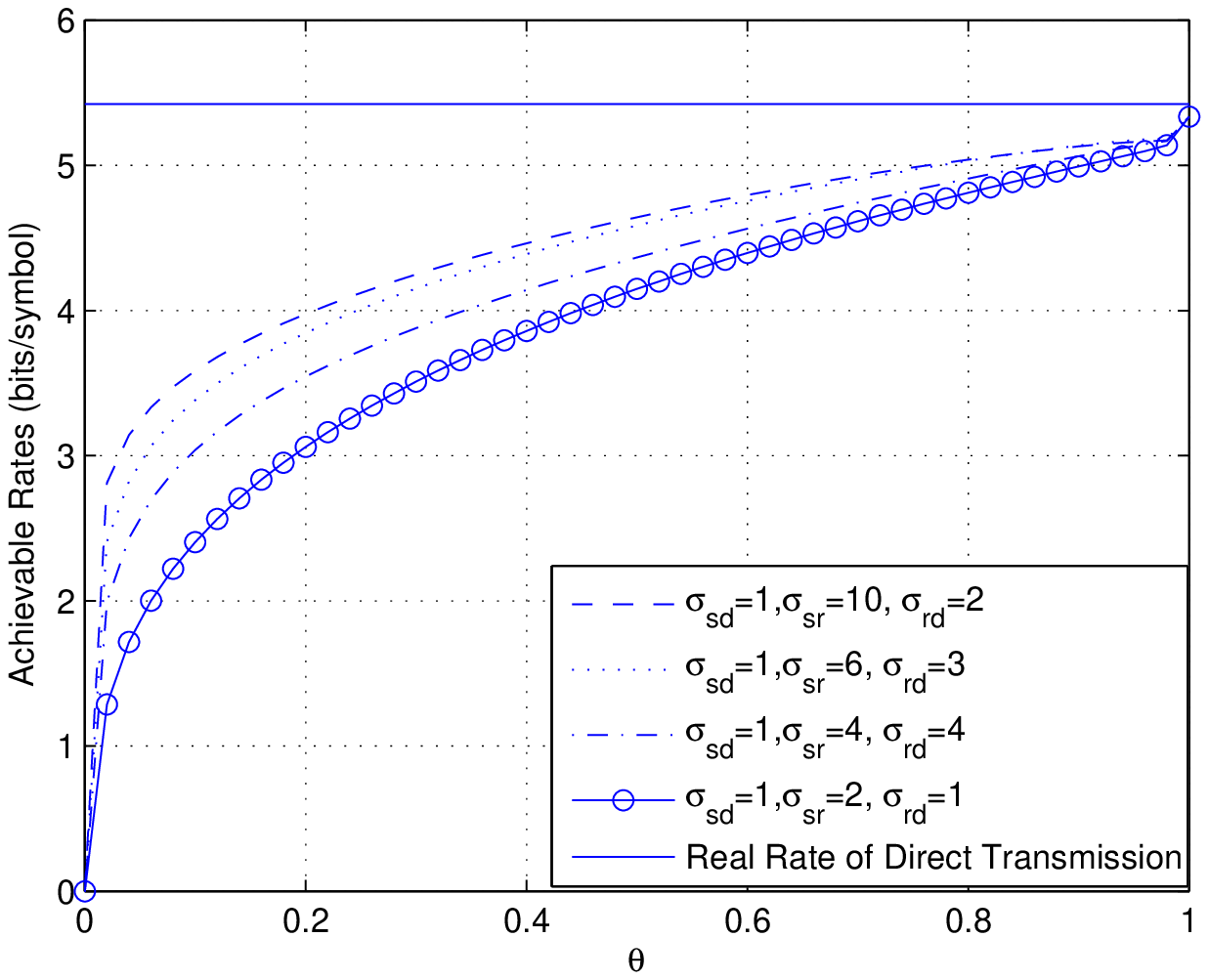}
\caption{Overlapped AF achievable rate vs. $\theta$. $P =100$,
$m=50$. } \label{fig:overaf}
\end{center}
\end{figure}

\begin{figure}
\begin{center}
\includegraphics[width = \figsize \textwidth]{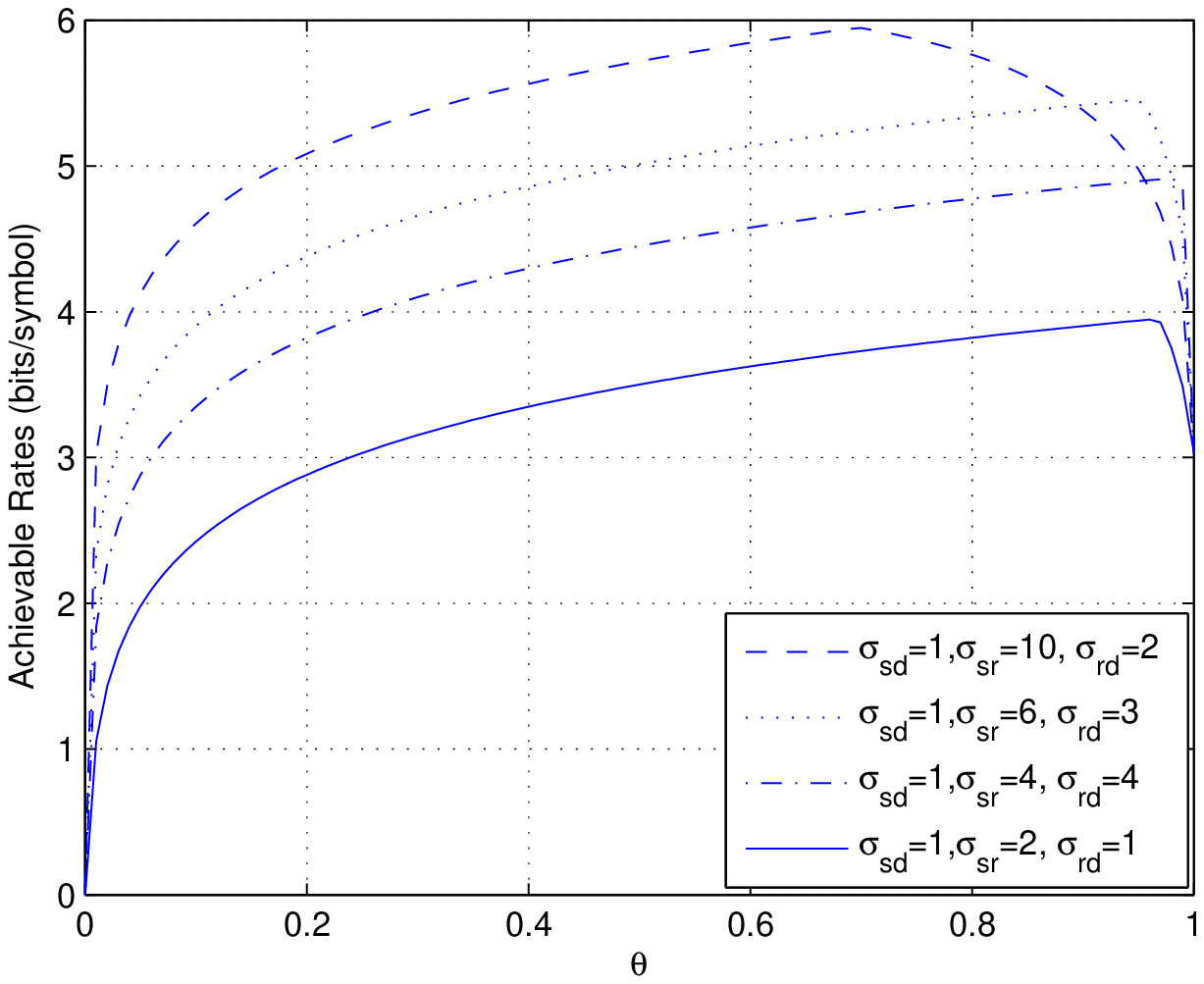}
\caption{Non-overlapped Parallel coding DF rate vs. $\theta$. $P
=100$, $m=50$. } \label{fig:4}
\end{center}
\end{figure}

In certain cases, source and relay are subject to a total power
constraint. Here, we introduce the power allocation coefficient
$\theta$, and total power constraint $P$. $P_s$ and $P_r$ have the
following relations: $P_s=\theta P$, $P_r=(1-\theta)P$, and
$P_s+P_r\leq P$. Next, we investigate how different values of
$\theta$, and hence different power allocation strategies, affect
the achievable rates. An analytical results for $\theta$ that
maximizes the achievable rates is difficult to obtain. Therefore, we
again resort to numerical analysis. In all numerical results, we
assume that $\alpha=0.5$ which provides the maximum of degree of
cooperation. First, we consider the AF. The fixed parameters we
choose are $P=100, N_0=1, \delta_s=0.1, \delta_r=0.1$. Fig.
\ref{fig:overaf} plots the achievable rates in the overlapped
transmission scenario as a function of $\theta$ for different
channel conditions, i.e., different values of
$\sigma_{sr},\sigma_{rd}, \text{ and }\sigma_{sd}$. We observe that
the best performance is achieved as $\theta \to 1$. Hence, even in
the overlapped scenario, all the power should be allocated to the
source and direct transmission should be preferred at these high SNR
levels. Note that if direct transmission is performed, there is no
need to learn the relay-destination channel. Since the time
allocated to the training for this channel should be allocated to
data transmission, the real rate of direct transmission is slightly
higher than the point that the cooperative rates converge as $\theta
\to 1$. For this reason, we also provide the direct transmission
rate separately in Fig. \ref{fig:overaf}. Further numerical analysis
has indicated that direct transmission over performs non-overlapped
AF, overlapped and non-overlapped DF with repetition coding as well
at this level of input power.
On the other hand, in Fig. \ref{fig:4} which plots the achievable
rates of non-overlapped DF with parallel coding as a function of
$\theta$, we observe that direct transmission rate, which is the
same as that given in Fig. \ref{fig:overaf}, is exceeded if
$\sigma_{sr} = 10$ and hence the source-relay channel is very
strong. The best performance is achieved when $\theta \approx 0.7$
and therefore $70\%$ of the power is allocated to the source.

\begin{figure}
\begin{center}
\includegraphics[width = \figsize\textwidth]{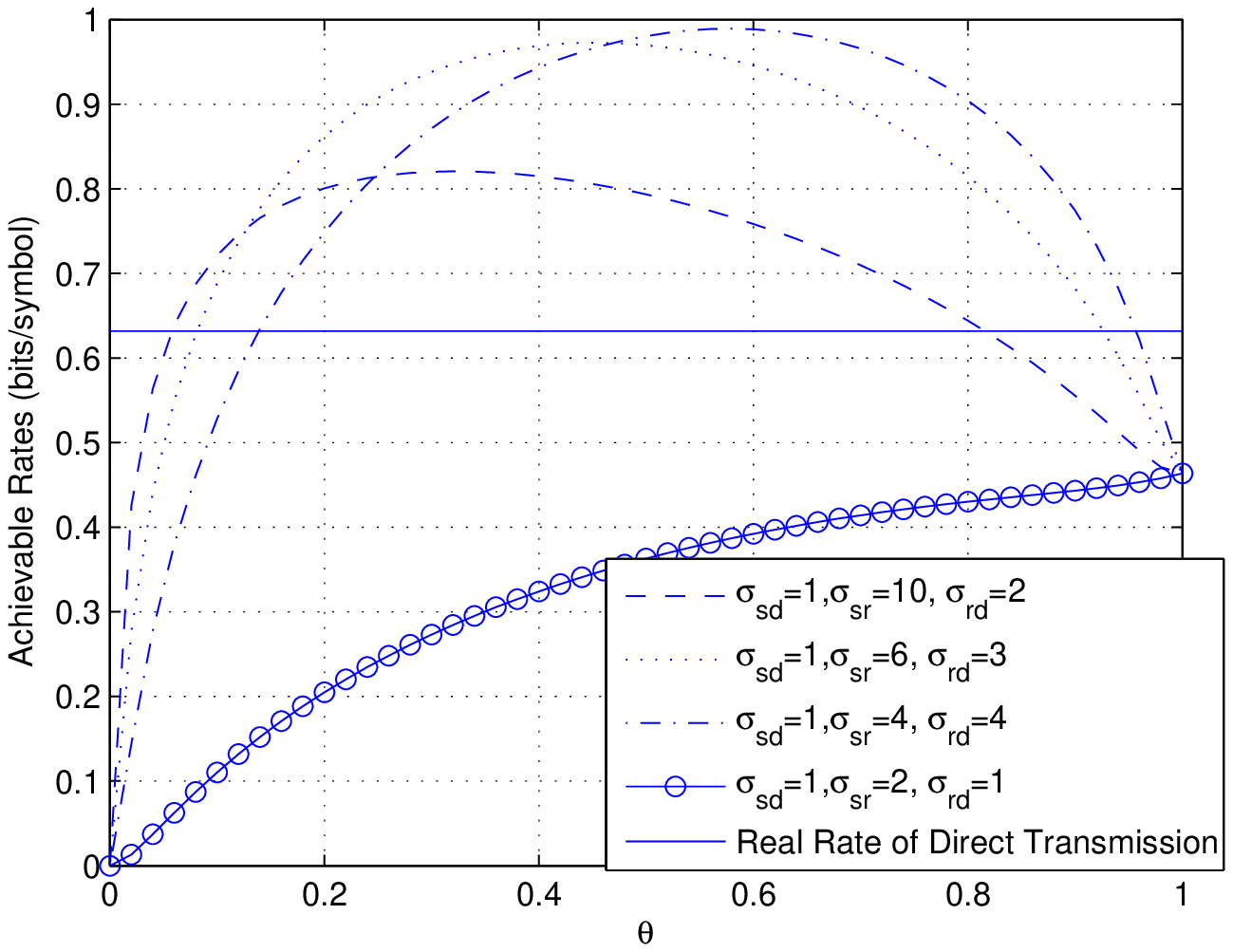}
\caption{Non-overlapped AF achievable rate vs. $\theta$. $P =1$,
$m=50$.} \label{fig:5}
\end{center}
\end{figure}
\begin{figure}
\begin{center}
\includegraphics[width = \figsize\textwidth]{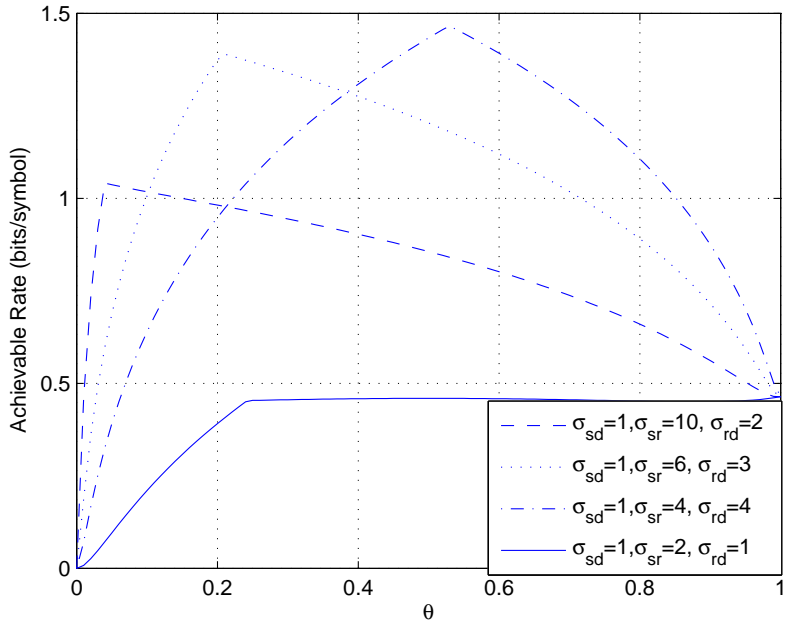}
\caption{Non-overlapped Repetition coding  DF rate vs. $\theta$. $P
=1$, $m=50$. } \label{fig:6}
\end{center}
\end{figure}
\begin{figure}
\begin{center}
\includegraphics[width = \figsize\textwidth]{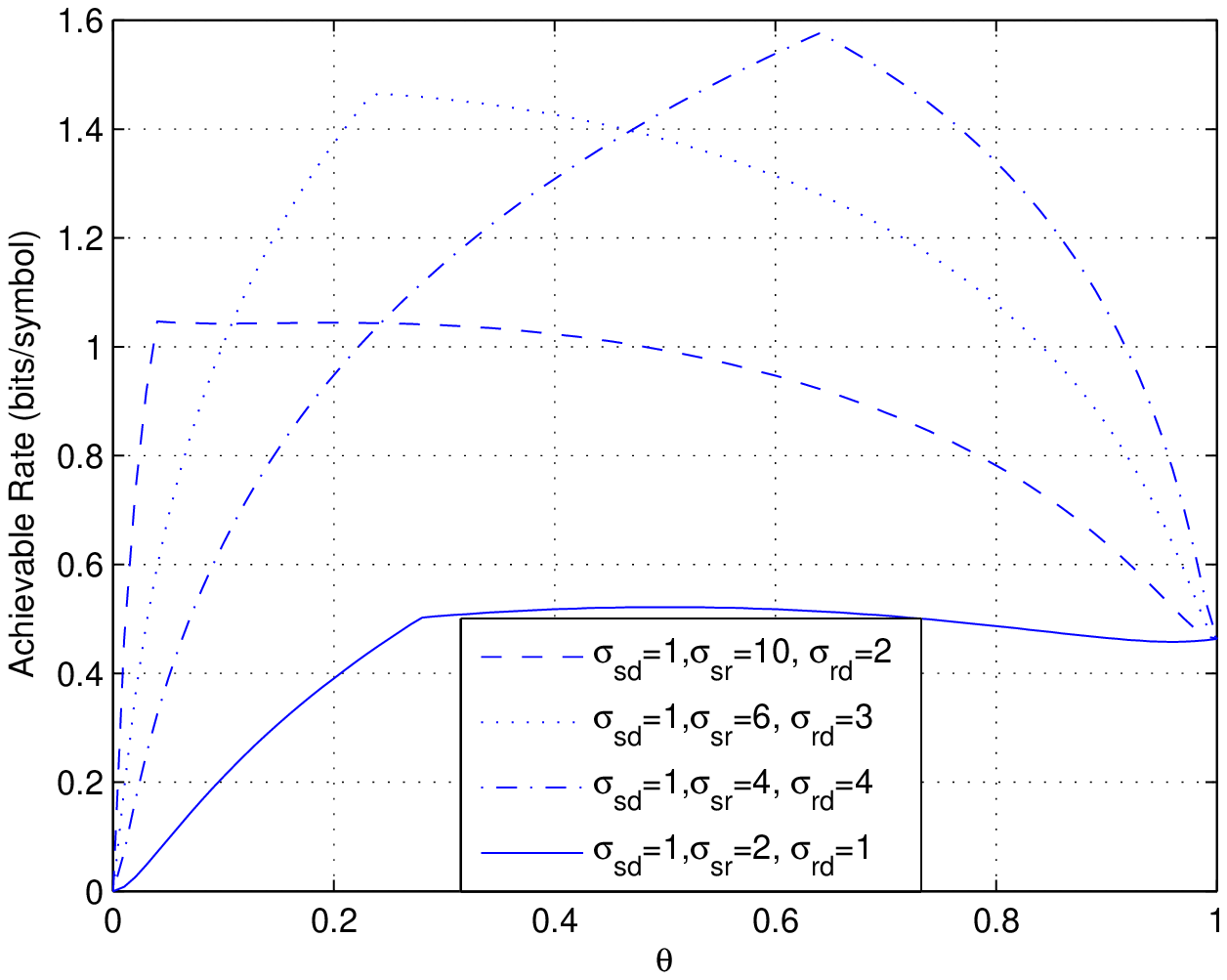}
\caption{Non-overlapped Parallel coding DF rate vs. $\theta$. $P
=1$, $m=50$. } \label{fig:7}
\end{center}
\end{figure}
\begin{figure}
\begin{center}
\includegraphics[width = \figsize\textwidth]{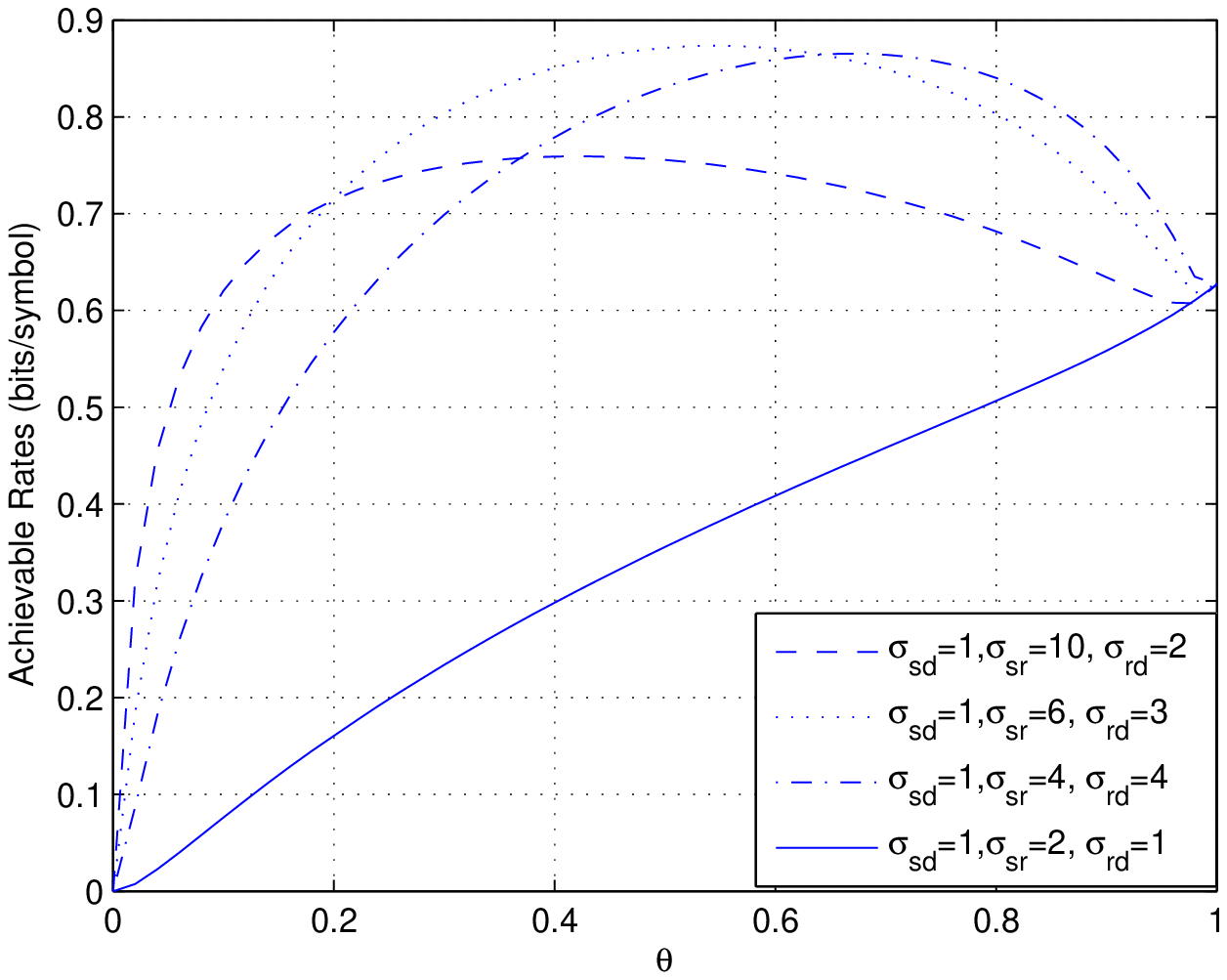}
\caption{ Overlapped AF achievable rate vs. $\theta$. $P =1$,
$m=50$.} \label{fig:AFoverlow}
\end{center}
\end{figure}

Figs. \ref{fig:5}, \ref{fig:6}, and \ref{fig:7} plot the
non-overlapped achievable rates when $P=1$. In all cases, we observe
that performance levels higher than that of direct transmission are
achieved unless the qualities of the source-relay and
relay-destination channels are comparable to that of the
source-destination channel (e.g., $\sigma_{sd}=1, \sigma_{sr}=2,
\sigma_{rd}=1$). Moreover, we note that the best performances are
attained when the source-relay and relay-destination channels are
both considerably better than the source-destination channel (i.e.,
when $\sigma_{sd} = 1, \sigma_{sr} = 4, \sigma_{rd} = 4$). As
expected, highest gains are obtained with parallel coding DF
although repetition coding incur only small losses.
Finally, Fig. \ref{fig:AFoverlow}  plot the achievable rates of
overlapped AF when $P=1$. Similar conclusions apply also here.
However, it is interesting to note that overlapped AF rates are
smaller than those achieved by non-overlapped AF. This behavior is
also observed when DF with repetition coding is considered. Note
that in non-overlapped transmission, source transmits in a shorter
duration of time with higher power. This signaling scheme provides
better performance as expected because it is well-known that flash
signaling achieves the capacity in the low-SNR regime in imperfectly
known channels \cite{Verdu}. 


\section{Energy efficiency} \label{sec:energyefficiency}

Our analysis has shown that cooperative relaying is generally
beneficial in the low-power regime, resulting in higher achievable
rates when compared to direct transmission. In this section, we
provide an energy efficiency perspective and remark that care should
also be taken when operating at very low $\tsnr$ values. The least
amount of energy required to send one information bit reliably is
given by\footnote{Note that $\frac{E_b}{N_0}$ is the bit energy
normalized by the noise power spectral level $N_0$.} $
\frac{E_b}{N_0} = \frac{\tsnr}{C(\tsnr)} $ where $C(\tsnr)$ is the
channel capacity in bits/symbol. In our setting, the capacity will
be replaced by the achievable rate expressions and hence the
resulting bit energy, denoted by
$\frac{E_{b,U}}{N_0}$, provides the least amount of normalized bit
energy values in the worst-case scenario and also serves as an upper
bound on the achievable bit energy levels in
the channel. 

We note that in finding the bit energy values, we assume that
$\tsnr=P/N_0$ where $P = P_r + P_s$ is the total power. The next
result provides the asymptotic behavior of the bit energy as $\tsnr$
decreases to zero.
\begin{lemma:bitenergylow} \label{lemma:bitenergylow}
The normalized bit energy in all relaying schemes grows without
bound as the signal-to-noise ratio decreases to zero, i.e.,
\begin{gather}
\left.\frac{E_{b,U}}{N_0}\right|_{I = 0} = \lim_{\tsnr \to 0}
\frac{\tsnr}{I(\tsnr)} = \frac{1}{\dot{I}(0)} = \infty.
\end{gather}
\end{lemma:bitenergylow}
\vspace{0.2cm} \emph{Proof}: The key point to prove this theorem  is
to show that when $\tsnr \to 0$, the mutual information decreases as
$\tsnr^2$, and hence $\dot{I}(0)=0$. This can be easily shown
because when $P \to 0$, in all the terms
$\frac{P_{s1}'|\hat{h}_{sd}|^2}{\sigma_{z_{d1}}^2}$,$\frac{P_{s1}'|\hat{h}_{sd}|^2}{\sigma_{z_{d2}}^2}$,
$\frac{P_{s1}'|\hat{h}_{sr}|^2}{\sigma_{z_{r}}^2},$
 $\frac{P_{r}'|\hat{h}_{rd}|^2}{\sigma_{z_{d}^r}^2}$,$\frac{P_{s2}'|\hat{h}_{sd}|^2}{\sigma_{z_{d1}}^2}$,$\frac{P_{s2}'|\hat{h}_{sd}|^2}{\sigma_{z_{d2}}^2}$,
$\frac{P_{s2}'|\hat{h}_{sr}|^2}{\sigma_{z_{r}}^2}$,$\frac{P_{s2}'|\hat{h}_{sd}|^2}{\sigma_{z_{d}^r}^2}$,
and $\frac{P_{r}'|\hat{h}_{rd}|^2}{\sigma_{z_{d}^r}^2}$ in Theorems
\ref{prop:asympcapaaa}-\ref{prop:asympcap}, the denominator goes to
a constant while the numerator decreases as $P^2$. Hence, these
terms diminish as $\tsnr^2$. Since $\log(1+x) = x + o(x)$ for small
$x$, we conclude that the achievable
rate expressions also decrease as $\tsnr^2$ as $\tsnr$ vanishes. 
\hfill $\square$

\begin{figure}
\begin{center}
\includegraphics[width = \figsize\textwidth]{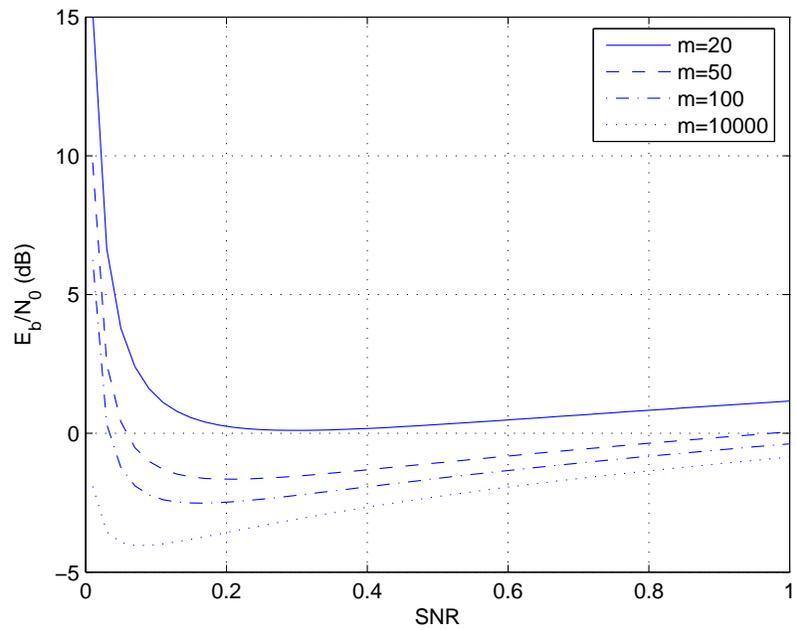}
\caption{ Non-overlapped AF $E_{b,U}/N_0$ vs. $\tsnr$}
\label{fig:nonoverAF}
\end{center}
\end{figure}

\begin{figure}
\begin{center}
\includegraphics[width = \figsize\textwidth]{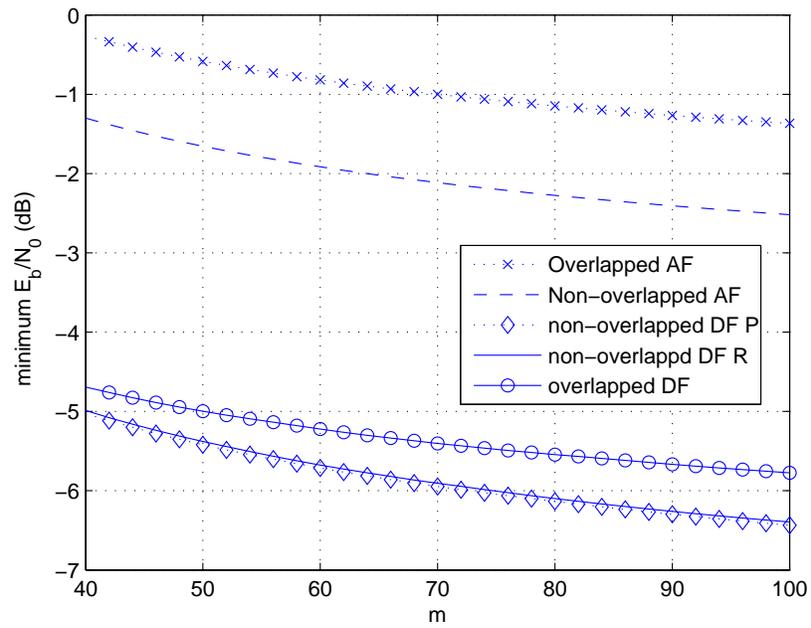}
\caption{ $E_{b,U}/N_0$ vs. $m$ for different transmission scheme}
\label{fig:enper}
\end{center}
\end{figure}

Theorem \ref{lemma:bitenergylow} indicates that it is extremely
energy-inefficient to operate at very low $\tsnr$ values. We
identify the most energy-efficient operating points in numerical
results. We choose the following numerical values for the fixed
parameters: $\delta_s = \delta_r = 0.1$, $\sigma_{sd}=1$,
$\sigma_{sr}=4$, $\sigma_{rd}= 4$, $\alpha = 0.5$, and $\theta=0.6$.
Fig. \ref{fig:nonoverAF} plots the bit energy curves as a function
of $\tsnr$ for different values of $m$ in the non-overlapped  AF
case. We can see from the figure that the minimum bit energy, which
is achieved at a nonzero value of $\tsnr$, decreases with increasing
$m$ and is achieved at a lower $\tsnr$ value. Fig. \ref {fig:enper}
shows the minimum bit energy for different relaying schemes with
overlapped or non-overlapped transmission techniques. We observe
that the minimum bit energy decreases with increasing $m$ in all
cases . We realize that DF is in general much more energy-efficient
than AF. Moreover, we note that employing non-overlapped rather than
overlapped transmission improves the energy efficiency. We further
remark that the performances of non-overlapped DF with repetition
coding and parallel coding are very close.


\section{Conclusion} \label{sec:conclusion}

In this paper, we have studied the imperfectly-known fading relay
channels. We have assumed that the source-destination, source-relay,
and relay-destination channels are not known by the corresponding
receivers a priori, and transmission starts with the training phase
in which the channel fading coefficients are learned with the
assistance of pilot symbols, albeit imperfectly. Hence, in this
setting, relaying increases the channel uncertainty in the system,
and there is increased estimation cost associated with cooperation.
We have investigated the performance of relaying by obtaining
achievable rates for AF and DF relaying schemes. We have considered
both non-overlapped and overlapped transmission scenarios. We have
controlled the degree of cooperation by varying the parameter
$\alpha$. We have identified the optimal resource allocation
strategies using the achievable rate expressions. We have observed
that if the source-relay channel quality is low, then cooperation is
not beneficial and direct transmission should be preferred at high
$\tsnr$s. On the other hand, we have seen that relaying generally
improves the performance at low $\tsnr$s. We have noted that DF with
parallel coding provides the highest rates. Additionally, under
total power constraints, we have identified the optimal power
allocation between the source and relay. We have again pointed out
that relaying degrades the performance at high $\tsnr$s unless DF
with parallel channel coding is used and the source-relay channel
quality is high. The benefits of relaying is again demonstrated at
low $\tsnr$s. We have noted that non-overlapped transmission is
superior compared to overlapped one in this regime. Finally, we have
considered the energy efficiency in the low-power regime, and proved
that the bit energy increases without bound as $\tsnr$ diminishes.
Hence, operation at very low $\tsnr$ levels should be avoided. From
the energy efficiency perspective, we have again observed that
non-overlapped transmission provides better performance than
overlapped transmission. We have also noted that DF is more energy
efficient than AF.

\end{spacing}

\newpage

%

%
%


\end{document}